\documentclass[11pt]{extarticle}
\usepackage{multicol}
\usepackage[]{graphicx}
\usepackage[]{xcolor}
\usepackage{alltt}
\usepackage[T1]{fontenc}
\usepackage[utf8]{inputenc}
\usepackage{multirow}
\usepackage{gensymb}
\usepackage{subfigure}
\usepackage{amsmath}
\usepackage{amssymb}
\usepackage{cite}
\usepackage{array}
\usepackage[]{authblk}
\usepackage{threeparttable}
\usepackage{textcomp}
\usepackage{dblfloatfix}
\usepackage{babel}
\usepackage[switch]{lineno}
\usepackage{grffile}

\usepackage{caption}
\usepackage{xr}
\makeatletter
\newcommand*{\addFileDependency}[1]{
  \typeout{(#1)}
  \@addtofilelist{#1}
  \IfFileExists{#1}{}{\typeout{No file #1.}}
}
\makeatother

\newcommand*{\myexternaldocument}[1]{
    \externaldocument{#1}
    \addFileDependency{#1.tex}
    \addFileDependency{#1.aux}
}
\myexternaldocument{supp}

\usepackage[colorlinks,
			linkcolor=blue,
			anchorcolor=green,
			citecolor=blue,
			urlcolor=blue
			]{hyperref}
\usepackage[nameinlink,noabbrev]{cleveref} % <-- new

\usepackage{fancyhdr}
\usepackage[a4paper,top=60pt,bottom=55pt,left=50pt,right=50pt]{geometry}

\usepackage{textcomp}

\def\keywords{\vspace{.3em}
{\textit{Keywords}:\,\relax%
}}
\def\endkeywords{\par}

\usepackage{float}
\floatstyle{plaintop}
\restylefloat{table}

\author[1,2,*]{Xiang Xu}
\author[1,*]{Xi Zhang}
\author[3]{Erik Bitzek}
\author[2]{Siegfried Schmauder}
\author[1]{Blazej Grabowski}
\affil[1]{Institute for Materials Science, University of Stuttgart, Pfaffenwaldring 55, 70569 Stuttgart, Germany}
\affil[2]{Institute for Materials Testing, Materials Science and Strength of Materials, University of Stuttgart, Pfaffenwaldring 32, 70569 Stuttgart, Germany}
\affil[3]{Computational Materials Design, Max Planck Institute for Sustainable Materials, Max-Planck-Straße 1, 40237 Düsseldorf, Germany.}
\affil[*]{\rm Emails: xiang.xu@imw.uni-stuttgart.de; xi.zhang@imw.uni-stuttgart.de}

\date{}
\title{Origin of the yield stress anomaly in L1$_2$ intermetallics unveiled with physically-informed machine-learning potentials}
\begin{document}
%TC:ignore
% \twocolumn[
% \begin{@twocolumnfalse}

\maketitle
\begin{sloppypar}
\vspace{-1cm}
\section*{Abstract}

The yield stress anomaly of L1$_2$ intermetallics such as Ni$_3$Al or Ni$_3$Ga is controlled by the so-called Kear-Wilsdorf lock (KWL), of which the formation and unlocking are governed by dislocation cross-slip.
Despite the importance of L1$_2$ intermetallics for strengthening Ni-based superalloys, microscopic understanding of the KWL is limited.
% leading to significant approximations in existing models for predicting the mechanical behavior of those intermetallics.
Here, molecular dynamics simulations are conducted by employing a dedicated machine-learning interatomic potential derived via physically-informed active-learning. The potential facilitates modelling of the dislocation behavior in Ni$_3$Al with near \textit{ab initio} accuracy. 
KWL formation and unlocking are observed and analyzed.
The unlocking stress demonstrates a pronounced temperature dependence, contradicting the assumptions of existing analytical models.
A phenomenological model is proposed to effectively describe the atomistic unlocking stresses and extrapolate them to the macroscopic scale. 
The model is general and applicable to other L1$_2$ intermetallics.
The acquired knowledge of KWLs provides a deeper understanding on the origin of the yield stress anomaly.

\vspace{0.15cm}
\noindent
\keywords
Yield stress anomaly; Dislocation cross-slip; L1$_2$ intermetallics; Molecular dynamics simulations; Machine-learning potentials.
\endkeywords

% \vspace{0.5cm}
% \end{@twocolumnfalse}
% ]
%TC:endignore

\section{Introduction}
Ni-based superalloys are used for turbine blades because they withstand thermal mechanical loadings under high turbine-entry temperatures~\cite{pollock2006nickel,long2018microstructural}. Over several generations of these superalloys and corresponding thermal barrier coatings, the turbine-entry temperatures have increased by 700\,K~\cite{reed2008superalloys}, significantly improving the thermodynamic efficiency of aircraft engines.
The outstanding thermal resistance mainly originates from a high volume fraction of L1$_2$-ordered precipitates.
In contrast to common structural materials, the yield stress of certain L1$_2$ intermetallics, e.g., Ni$_3$Al~\cite{golberg1998effect} or Ni$_3$Ga~\cite{takeuchi1973temperature}, increases with temperature, typically accompanied with an anomalously increasing work-hardening rate.
As this so-called yield stress anomaly (YSA) is pivotal for strengthening advanced alloys, the steady increase of understanding YSA has been a key ingredient not only to the evolution of Ni-based superalloys~\cite{pollock2006nickel,wang2017understanding,long2018microstructural} and Co-based superalloys~\cite{eggeler2021precipitate,suzuki2015l12,chen2022improving}, and also for the development of L1$_2$ strengthened high-entropy alloys~\cite{cao2021novel,yang2018l12}.

However, the origin of YSA is still not satisfactorily clarified~\cite{wang2017understanding}.
What is known from transmission electron microscopy (TEM) on samples deformed in the temperature region of the YSA~\cite{veyssiere1985presence,sun1988tem} is that the dislocations in Ni$_3$Al exhibit a unique non-planar dislocation core structure---nowadays referred to as the Kear-Wilsdorf lock (KWL).
The dislocation core was shown to evolve through cross-slip~\cite{molenat1991dislocation,coupeau2020atomic} in which three planar defects are involved: two antiphase boundaries (APBs) on the (100) and (111) planes plus a complex stacking fault (CSF). 

Several analytical models~\cite{paidar1984theory,yoo1986theory,hirsch1992new,caillard1996model1,kruml2002dislocation,choi2007modelling,demura2007athermal} have been proposed to comprehend KWLs and their relation to YSA, considering factors like the difference between the formation energies of the (111)APB and (100)APB~\cite{paidar1984theory}, and torque interactions between the superpartials~\cite{yoo1986theory}. 
The ``APB-jump'' phenomenon observed in \textit{in situ} TEM experiments~\cite{molenat1991dislocation} brought forward a model based on the competition between the formation and unlocking of the incomplete KWL~\cite{caillard1996model1}.
An incomplete KWL is built up from APBs on both the (100) and (111) planes while a complete KWL contains only an APB on the (100) plane (see Figure~\ref{sfig_KWL} in Supplementary Material).
A common limitation of existing models is the assumption of an athermal unlocking process~\cite{caillard1996model1,choi2007modelling}.

Despite the importance of KWLs, the understanding of their formation and unlocking is limited, especially regarding the atomistic evolution during cross-slip. 
A close atomistic inspection is difficult with experiments, but becomes feasible with atomistic simulations. 
For example, the embedded atom method (EAM) has been used to investigate the dissociation of superdislocation in Ni$_3$Al~\cite{yoo1989atomistic,wen1997effect}, to calculate the nucleation energy of the cross-slip process for a single dislocation~\cite{parthasarathy1996atomistic,ngan2004atomistic} and for intersecting dislocations~\cite{rao2012atomistic}.
These simulation studies focused on energetics at 0\,K and thus neglected entropy contributions relevant at elevated temperatures.
Recently, the temperature-dependent dislocation dynamics of edge dislocation in Ni$_3$Al has been investigated using EAM potential~\cite{wakeda2023atomistic}.
However, atomistic simulations for the behavior of screw superdislocations, particularly when related to KWLs and cross-slip processes at elevated temperatures, are still absent.
This is presumably due to the limited accuracy of existing potentials in predicting the energetics of planar defects, which prevents the occurrence of dislocation cross-slip.
%Therefore, a highly accurate interatomic potential is demanded for revealing the detailed dislocation behavior in the atomistic level.
% Another drawback is the limited accuracy of the classical interatomic potentials which, for example, prevents the expected formation of KWLs. 

In the present study, a machine-learning interatomic potential (MLIP), specifically a Moment Tensor Potential (MTP)~\cite{shapeev2016moment,gubaev2019accelerating}, is developed and utilized to simulate the dislocation activity in Ni$_3$Al at near \textit{ab initio} accuracy and with inclusion of finite temperature effects. The most critical aspect in designing MLIPs is the proper choice of the fitting dataset from density functional theory (DFT). To guarantee an accurate description of the dislocation, we utilize a physically-informed active-learning scheme and evaluate the MLIP specifically on the temperature dependence of the planar defect energies.
The MLIP enables the study of the unlocking of KWLs by shearing at different temperatures in molecular dynamics (MD) simulations.
Based on the obtained data, a phenomenological model for \emph{thermally} activated KWL unlocking is derived.

%%%%%%%%%%%%%%%%%%%%%%%%%%%%%%%%%%%% MTP
\section{Results}
\subsection{Physically-Informed Active Learning}

\begin{figure*}[!htb]
\centering
\includegraphics[width=\textwidth]{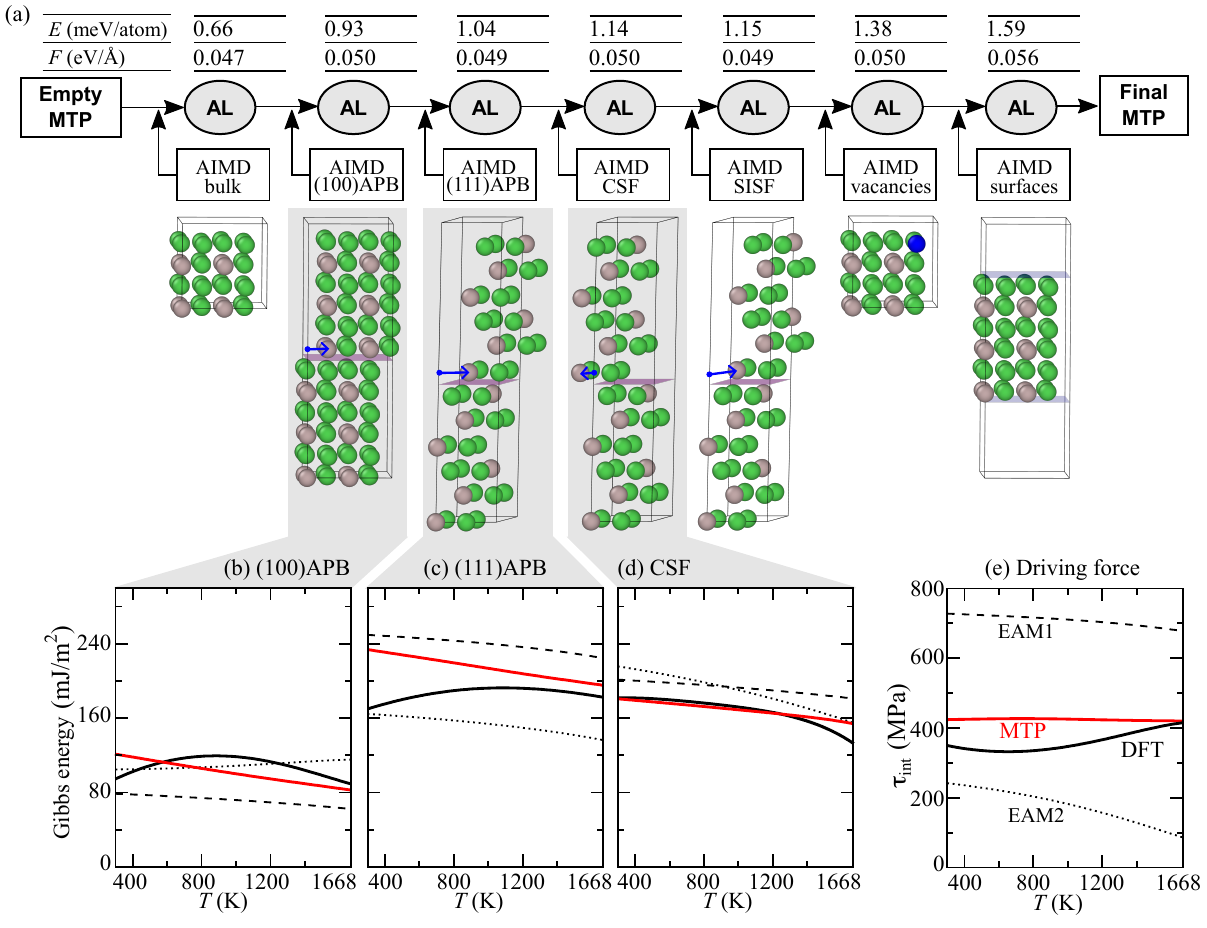}
\caption{Machine-learning potential design. (a) Flow-diagram of the proposed physically-informed active-learning scheme.
Every oval labelled ``AL'' represents a standard active-learning step for the indicated geometry, i.e., bulk structure, (100)APB, (111)APB, CSF, superlattice intrinsic stacking fault (SISF), Ni-vacancy, Al-vacancy, and a surface with its normal along the [100] direction.
The atomic structures are simplified for a better illustration. The blue arrows indicate the vectors of the relative shifts needed to generate the planar defects.
The input for each subsequent AL-oval is 1) the MTP from the previous AL-step  and 2) a set of AIMD configurations for the new geometry to provide basic structural information. 
The root mean square errors (RMSEs) above each oval represent the fitting errors when completing the respective AL-step.
(b) - (d) Temperature-dependent Gibbs formation energies  of the planar defects predicted by the \emph{final} MTP.
(e) Driving force for the cross-slip process originating from the difference of the two APB energies in combination with the anisotropic factor (cf. Equation~\eqref{seq_CS_form} and the related discussion in Supplementary Material).
DFT results are from previous works~\cite{xu2023APBs,xu2023CSF}.
Dashed (EAM1) and dotted (EAM2) curves show results from the EAM potentials modified by Mishin \textit{et al.}~\cite{mishin2004atomistic} and Du \textit{et al.}~\cite{du2012construction}, respectively.
}
\label{fig_AL}
\end{figure*}

Active-learning (AL)~\cite{gubaev2019accelerating} has been successfully used in different machine-learning studies~\cite{lee2023atomic,yin2021atomistic}. However, for the present purpose, the standard AL scheme cannot be applied due to the large, DFT-inaccessible supercell size required to model a KWL ($>$ 1 million of atoms). Hence, we devise a modified AL scheme, in the spirit of 
% a very recent scheme applied to large-scale silicon-oxygen systems~\cite{erhard2024modelling}. 
two very recent schemes applied to large-scale silicon-oxygen systems~\cite{erhard2024modelling} and to dissociated partial dislocations~\cite{mismetti2024automated}.
The key idea is to decompose the KWL into its physically relevant parts which \emph{can} be modelled with (periodic) DFT. The choice of the relevant parts is guided by domain expertise and is displayed in Figure~\ref{fig_AL}(a). Besides the perfect bulk, which is clearly required as the basic fitting input, we know that a KWL is formed by the (100)APB, the (111)APB, and the CSF. For the present shearing simulations, we also need the (111) surface due to the boundary conditions. Additionally, we include the superlattice intrinsic stacking fault (SISF), and Al and Ni vacancies into the fitting dataset for a broader applicability of the MLIP, for example for investigations of creep properties.

For each such geometry a usual AL is performed at a high temperature (here 1600 K). The different AL steps are executed successively in a row. We start with the perfect bulk AL and take the resulting MTP as input to the next AL step for the (100)APB. This process continues for the remaining geometries until a final MTP is obtained. The sequence is displayed in Figure~\ref{fig_AL}(a). Fitting the MTP in such a systematic way allows us to monitor the accuracy at each step. As usual, one measure of the MTP accuracy is the root mean square error (RMSE) in energies and forces. The force RMSE stays almost unchanged while the energy RMSE increases by a factor of 2.4 during the AL steps (cf.~Figure~\ref{fig_AL}(a)). The final training errors of $1.59$~meV/atom and 0.056~eV/\AA\ are small considering the high temperature and structural complexity of the fitting dataset.

An additional measure that we employ here and that gives us control over the accuracy in describing the geometry of interest, i.e., the KWL, is the locally resolved extrapolation grade. The extrapolation grade is a metric that quantifies how far away a certain atomic configuration is located in phase space with respect to the fitting dataset.\footnote{Numerically, extrapolation
by the MTP occurs if the grade is greater than unity. The higher the extrapolation grade the severer is the extrapolation
by the MTP.} The advantage is that no extra DFT calculation is required and thus the grade can be computed for the target geometry even if a large supercell is required, as for the KWL. Specifically, we utilize our final MTP to generate KWL snapshots for which we resolve the extrapolation grade locally. We do this for the different MTPs obtained from our physically-informed AL scheme and investigate how the grade changes along the sequence of AL steps (Supplement Figure~\ref{sfig_localGrade}). The MTP trained with bulk structures only (MTP-bulk) describes the perfect bulk atoms (i.e., 98.8\% of all atoms) with a grade of $\lesssim$\,$1$ which indicates an interpolative behavior. Atoms belonging to the KWL are revealed to have a slightly higher grade which indicates an extrapolative behavior. Upon inclusion of the (100)APB and (111)APB geometries into the physically-informed AL, the extrapolation grade steadily improves, reaching a grade of $\lesssim$\,$1$ also in the KWL region. The analysis of the local grade thus gives increased confidence in the predictive capability of the MTP in the most relevant simulation region.

% Notably, adding a new category of structures changes marginally the training errors, for example from 0.51 to 0.77~meV/atom (no change on the force error) when adding (100)APB and (111)APB as shown in Figure~\ref{fig_AL}(a), indicating the stability of MTP.
% The final training error of energy $1.24$~meV/atom and force 0.025~eV/\AA~demonstrates the high accuracy of the trained MTP.
%Remarkably, the resulting low extrapolation grades for superdislocation and cross-slipped configurations, which are not included in the training set, convincingly demonstrate the utilized training set is representative and ensure the reliability of MTP to predict the complex dislocation motion.
As a further quality measure, the final MTP is evaluated on the free energy surfaces for bulk Ni$_3$Al, the APBs, and the CSF, up to the melting point 1668\,K. The thermal properties of perfect bulk Ni$_3$Al are well predicted by the MTP over a large temperature range (Figure~\ref{sfig_ThermProp} in the Supplementary Material). The resulting temperature-dependent planar defect energies (red curves in Figure~\ref{fig_AL} from (b)-(d)) are likewise in good agreement with the explicit DFT results (black curves) \cite{xu2023APBs,xu2023CSF}, in particular in the most relevant temperature range for the KWL simulations around 1000\,K. Of special importance is the fact that the internal driving force to form a KWL (Figure~\ref{fig_AL}(e)), which originates from the difference of the two APB energies in combination with the anisotropic factor (cf.~Equation~\eqref{seq_CS_form}) and which the two available EAM potentials strongly under- or overestimate, is close to DFT for the MTP.

% As shown in Figure~\ref{fig_AL} from (b)-(d), the predicted thermal properties are in good agreement with experiment and density functional theory (DFT).
% The deviations, e.g., in the isobaric heat capacity and the thermal
% expansion coefficient, between DFT and MTP originate from thermal electronic excitations. %which cannot be captured by the current version of the MTP package.
% For the temperature region of interest (above 300\,K), the utilized MTP well reproduces the DFT energetics of the planar defects~\cite{xu2023APBs,xu2023CSF}.
% In the temperature region of interest (above 300\,K), the utilized MTP well predicts the energetics of planar defects comparing with DFT results~\cite{xu2023APBs,xu2023CSF}.
% It should be mentioned that the deviation, for example for (111)APB at the low temperature region, will not change the mechanism behind the dislocation behavior but moderately adjust the investigated stresses.
% Additionally, the thermal properties of Ni$_3$Al are very well predicted by this MTP, as shown in Figure~\ref{sfig_ThermProp} in the Supplementary Material.

%%%%%%%%%%%%%%%%%%%%%%%%%%%%%%%%%%%%%%%%%%%%%%%%%%%%%%%%

\subsection{Formation of KWL}\label{sec_CS}

\begin{figure*}[!htb]
\centering
\includegraphics[width=\textwidth]{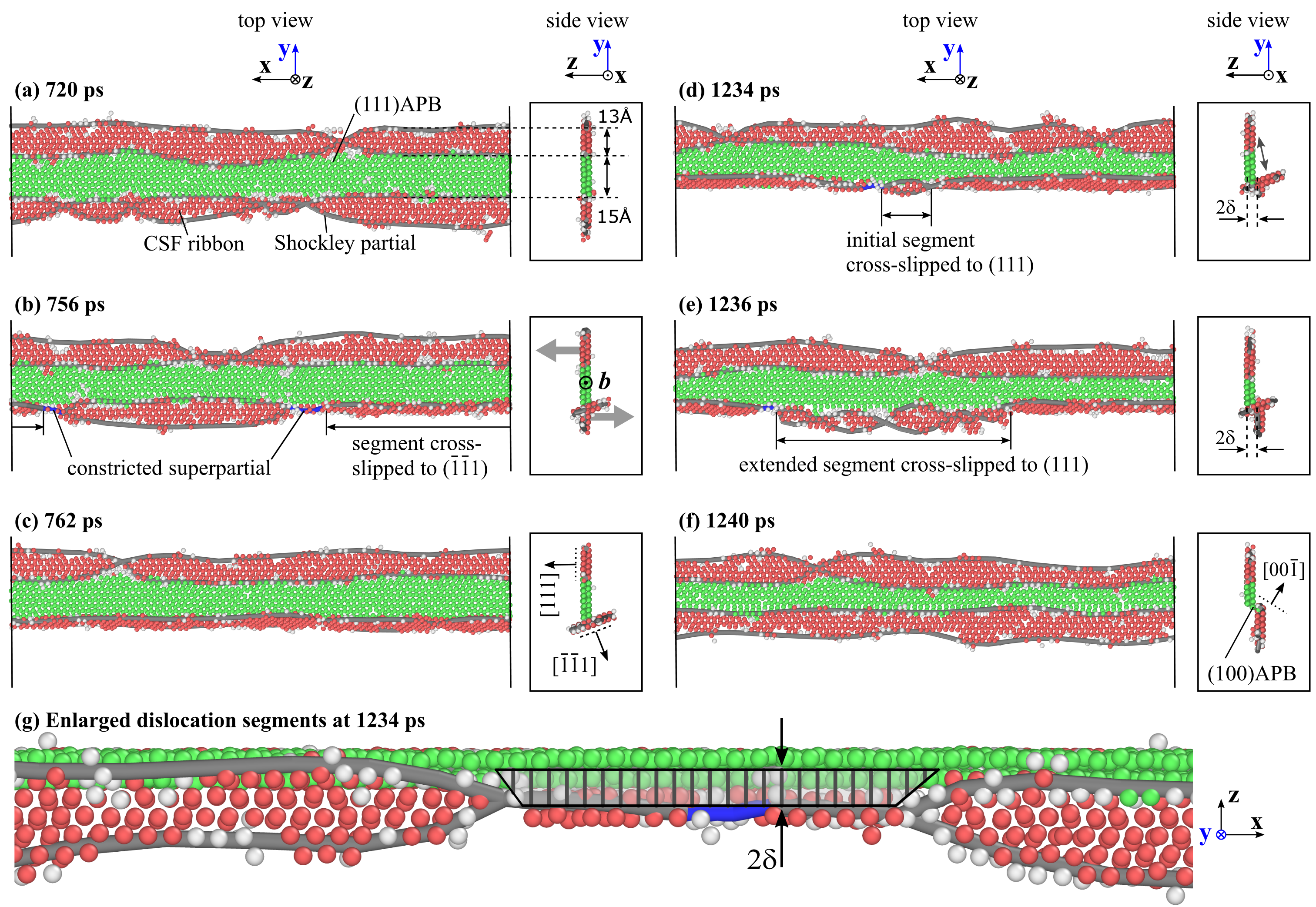}
\caption{Formation of an incomplete KWL at 980\,K. 
Snapshots (a) to (c) display the first cross-slip from a (111) to a ($\bar{1}\bar{1}1$) plane, and snapshots (d) to (f) the second cross-slip from the ($\bar{1}\bar{1}1$) plane to another (111) plane. 
(g) Enlarged dislocation segments at 1234\,ps.
The dislocation structure is analyzed by using the Dislocation Extraction Algorithm (DXA) in Ovito~\cite{stukowski2010extracting}.
Green and red atoms highlight the APB region and the CSF ribbon, respectively. %, while gray atoms are in amorphous structure.
For a clear illustration, atoms not related to the dislocation core structure are removed.
The gray tubes highlight Shockley partials and the blue ones constricted superpartials.
% The projected cross-slip distance is $2\delta_{\{111\}}$, twice the distance between atomic planes along $[111]$ direction.
}
\label{fig_CS}
\end{figure*}

With the optimized MTP, the KWL can be readily created.
Figure~\ref{fig_CS} documents the formation of an incomplete KWL as observed in an MD simulation at 980 K (without external loading). 
The corresponding dislocation core configurations are detailed in Section~\ref{core_conf} in the Supplementary Materials.
In additional test simulations with EAM potentials~\cite{mishin2004atomistic,du2012construction}, such a spontaneous KWL formation is not observed at similar temperatures.\footnote{Our calculations show that for the Mishin-EAM~\cite{mishin2004atomistic} a KWL forms for an overheated system at 1200 K and that no KWL formation is feasible for the Du-EAM~\cite{du2012construction} within the typical MD time scale.}
%With the trained MTP, an incomplete KWL can be readily created.
%Figure~\ref{fig_CS} shows two cross-slip processes during equilibration of a superdislocation at 980\,K without external loading. 

Snapshots (a) to (c) illustrate the first cross-slip process required to form a KWL.
In snapshot (a), the dislocation is initially fully dissociated on the (111) plane, with each superpartial split into a pair of Shockley partials separated by a CSF ribbon. Then, the lower superpartial cross-slips from the (111) plane to the $(\bar{1}\bar{1}1)$ plane, partially in (b) and fully in (c).
At 1234\,ps, the second cross-slip of the lower superpartial, from the $(\bar{1}\bar{1}1)$ plane to another (111) plane, is initiated (Figure~\ref{fig_CS}(d)).
When this double cross-slip process is finished, an incomplete KWL has formed (Figure~\ref{fig_CS}(f)).

It is worth noting that the distance between the original (111) plane and the (111) plane onto which the superpartial moves during the second cross-slip process is $2\delta$ (see the side view in Figure~\ref{fig_CS}(d) and (e) and the enlarged front view in Figure~\ref{fig_CS}(g)), where $\delta$ labels the distance between the two nearest atomic planes along the $[111]$ direction.
Our results are in agreement with a recent \textit{in situ} scanning tunnelling microscopy study~\cite{coupeau2020atomic} which reported $2\delta$ at room temperature and $m\delta$ (with $m\geq 2$) at higher temperatures in Ni$_3$Al (1 at.\% Ta), however, without an interpretation.
It is the strong internal forces that drive the CSF ribbon in the lower superpartial away from the original (111) plane. An internal torque due to the anisotropic elastic interaction between the two screw superpartials~\cite{yoo1986theory,yoo1987stability}, indicated in the side view of Figure~\ref{fig_CS}(b) by the gray arrows, pushes the upper superpartial to the left and the lower superpartial to the right (note that the Burgers vector points out of the paper plane). In the analyzed simulation, the lower superpartial cross-slips first. (The leading and trailing superpartials have the same probability to cross-slip onto a (100) plane, as indicated by Figure~\ref{sfig_KWL_1100K} in the Supplementary Materials.) When it is located on the ($\bar{1}\bar{1}1$) plane, the repulsive force between the two superpartials (cf.~double headed arrow in the side view in Figure~\ref{fig_CS}(d)) drives the lower superpartial to cross-slip again, now to the (111) plane which is 2$\delta$ away from the original one. The observed atomistic shape of the cross-slipped segment, specifically the well-defined 2$\delta$ distance over the full segment shown in Figure~\ref{fig_CS}(g), supports the previously suggested ``double-jog'' mechanism \cite{hirsch1992new}. 
MD observations show that the jogs are highly mobile and expand rapidly along the dislocation line.
Incomplete KWLs with $3\delta$ or $4\delta$ likewise occur in the MD simulations. They form by consecutive cross-slips of both of the superpartials.

The present MD simulations reveal the importance of the superpartial splitting and of the corresponding CSF on the cross-slip behavior. This is a crucial insight, since available phenomenological models~\cite{paidar1984theory,hirsch1992new,caillard1996model1} do not explicitly take into account the screw superpartial splitting into Shockley partials.
This approximation of considering only \emph{constricted} superpartials favors $1\delta$-KWL formation at low temperatures and renders $1\delta$-KWLs to be the intermediate state for forming $m\delta$-KWLs at elevated temperatures.
In contrast, the MD simulations show that a $2\delta$-KWL can form directly without an intermediate $1\delta$-KWL.
We conclude that this is due to the CSF ribbon spanning on the $(\bar 1 \bar 1 1)$ plane.
This finding highlights the necessity to carefully consider the effect of the superpartial splitting into Shockley partials and the CSF ribbon on the dislocation behavior.

\subsection{Unlocking of KWL\label{sec:unlock}}

\begin{figure*}[!htb]
\centering
\includegraphics[width=\textwidth]{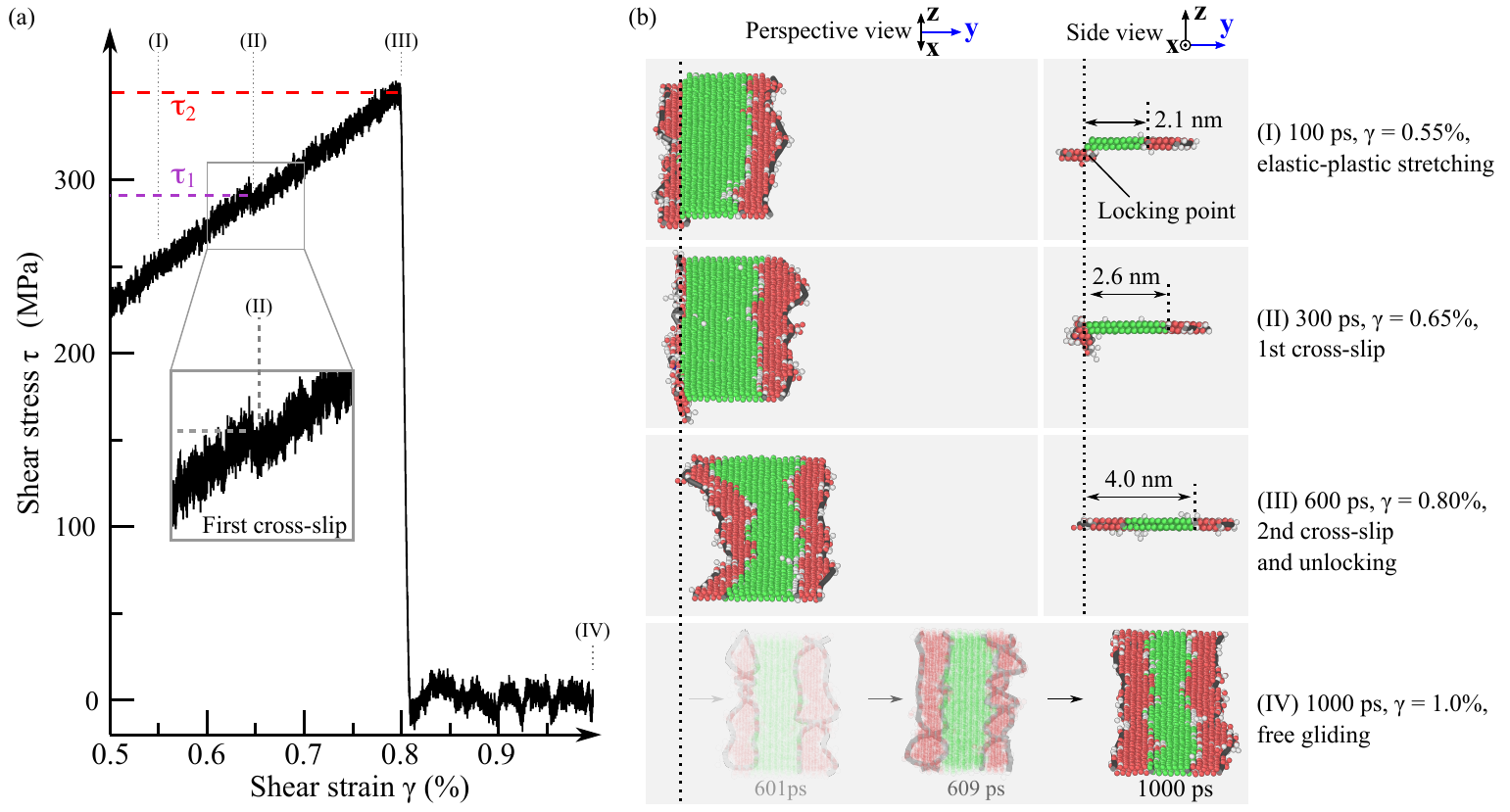}
\caption{Unlocking of a KWL at $T=1000$\,K and $\dot{\gamma} =  5\times10^{-6}~\mathrm{ps}^{-1}$.
(a) Corresponding stress-strain curve and
(b) snapshots at representative times highlighting the evolution of the KWL during unlocking.
Snapshot (IV) has only a perspective view. The given times are referenced with respect to the time at the shear strain of $\gamma=0.5\%$.
Atom colors have the same meaning as in Figure~\ref{fig_CS}.
The moving direction, i.e., the $y$-axis, is colored in blue.
}
\label{fig_un-IKWL-1000K}
\end{figure*}

The above-discussed KWL with the cross-slip distance of $2\delta$ was sheared at different temperatures $T$ and shear rates $\dot \gamma$.
The unlocking of the KWL at temperatures $\lesssim$\,$1000$\,K involves a two-step cross-slip process, in reverse order compared to its formation.
Figure~\ref{fig_un-IKWL-1000K} shows a representative stress-strain curve %at $T=1000$\,K and $\dot{\gamma}=5\times10^{-6}$~ps$^{-1}$ 
in (a) and selected dislocation structures at critical simulation times in (b).

The shear stress increases with strain at a slope of $45$\,GPa until point (III) when full unlocking occurs.
At the earlier point (II), the trailing superpartial cross-slips from its nesting (111) plane to the $(\bar{1}\bar{1}1)$ plane (difference between the configurations in (I) and (II) in Figure~\ref{fig_un-IKWL-1000K}(b)) accompanied by a small stress drop. 
The second cross-slip of the trailing superpartial from the $(\bar{1}\bar{1}1)$ plane to the (111) glide plane, which hosts the leading superpartial and the APB, happens at point (III) and is followed by a rapid glide of the dislocation (cf.~perspective view (IV) in Figure~\ref{fig_un-IKWL-1000K}(b)).
Accordingly, a stress collapse is observed in the stress-strain curve.
After unlocking, the released  dislocation either glides freely on the (111) plane or cross-slips to a (100) plane forming a KWL again.
The latter occurs in particular at higher temperatures.

\begin{figure*}[!htb]
\centering
\includegraphics[width=\textwidth]{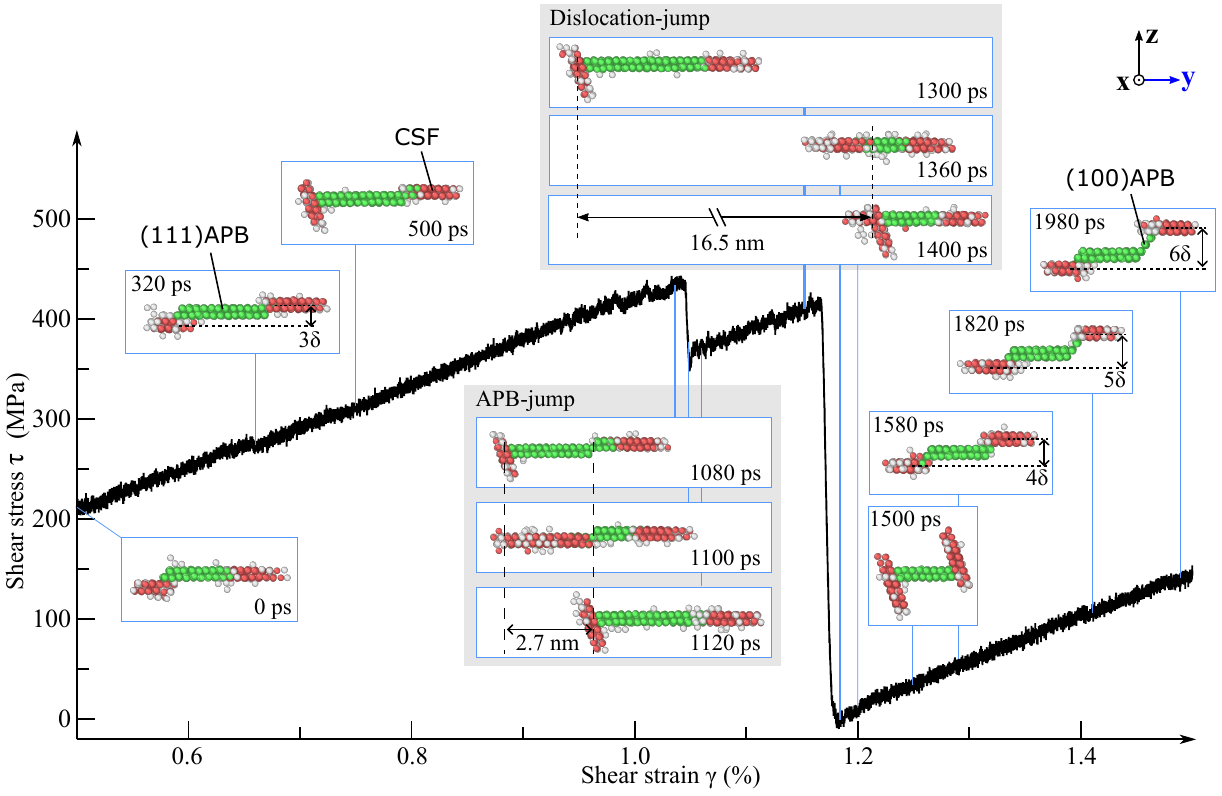}
\caption{Unlocking of a KWL at $T=1050$\,K and $\dot{\gamma}=5\times10^{-6}~\mathrm{ps}^{-1}$.
The given times are referenced with respect to the time at $\gamma=0.5\%$.
Atom colors have the same meaning as in Figure~\ref{fig_CS}.
}
\label{fig_un-IKWL-1050K}
\end{figure*}

An exemplary run at a slightly higher temperature of 1050\,K is shown in Figure~\ref{fig_un-IKWL-1050K}. Many cross-slip events occur during this simulation and various distinct core configurations of the superdislocation are observed.
In addition to the $2\delta$ configuration found for the KWL-formation and unlocking simulation at lower temperatures, here, core configurations can also exist with larger $\delta$ values for the superpartials separation, e.g., $3\delta$ (snapshot at 320\,ps), $4\delta$ (1580\,ps), $5\delta$ (1820\,ps), or $6\delta$ (1980\,ps). In contrast to the initial KWL-formation, the double cross-slip events at larger separations occur in steps of $1\delta$, which is a consequence of a smaller driving force. Some of the cross-slip events can be correlated with previous TEM observations~\cite{molenat1991dislocation}.

The ``APB-jump'' observed in \textit{in situ} TEM experiments~\cite{molenat1991dislocation} is replicated on the atomic scale at a shear strain of 1.05\%, as illustrated in Figure~\ref{fig_un-IKWL-1050K} in the box labelled ``APB-jump''.
This kind of jump starts from a core structure with two (111)APBs lying connected on two neighboring (111) planes. During the APB-jump, the trailing superpartial cross-slips from the ($\bar{1}\bar{1}1$) to the lower (111) plane and glides on this (111) plane by a short distance (2.7 nm), which reflects the (111)APB width corresponding to the instantaneous loading state, before cross-slipping to another ($\bar{1}\bar{1}1$) plane. %APB-jumps in the MD simulations occur for different initial geometries, i.e., for different spacings (1$\delta$, 2$\delta$, or 3$\delta$) between the two initially occupied (111)APB planes.
% and its exact jump distance, of 2.7\,nm in this work, is related to the (111)APB energy and the stress state before jump.

Another kind of jump (labelled ``dislocation-jump'') happens at a slightly higher strain of 1.17\%,
during which the released dislocation moves shortly freely on the (111) plane but quickly gets stopped due to a cross-slip event. 
This ``dislocation-jump'' is, from a broader perspective, similar to the previously observed ``long distance jump''~\cite{molenat1991dislocation} but different in the details of the mechanism proposed in that work.
From the MD snapshot at 1400\,ps, it is the trailing superpartial that firstly cross-slips to another $\{111\}$ plane rather than the leading one.
% The jump distance depends on few factors that can affect its gliding on a (111) plane, e.g., temperature, CSF energy, dislocation length etc.
A more detailed analysis of the MD simulations shows the mechanism of this ``dislocation-jump'' to be an inertia effect which asymmetrically impacts the motion of the leading and trailing superpartials. The consequence are asymmetric elastic vibrations of the core structure on the slip plane, different velocities of the leading and trailing superpartial, and eventually different cross-slip rates. 
We speculate that this dislocation jump is related to the jerky movement of dislocations in L1$_2$ intermetallics and suggest further research in this direction.
% (Note that a ``dislocation-jump'' with the leading superpartial cross-slipping to (100) plane was also observed in other shear simulations executed in this work.)
% This jump has also been observed in shear simulations at different temperatures with different jump distances.

% Notably, almost every cross-slip and the associated driving force can be recognized by change in stress.
% In general, a cross-slip driven by the internal interaction increases the stress, e.g., the one at 1400\,ps, while a cross-slip by the external loading decreases the stress.
% The one occurs at 1580\,ps, as an exception, shows no visible change in the shear stress. This is due to a compensation between the downward cross-slip of the trailing superpartial (driven by the internal interaction) and the upward cross-slip of the leading superpartial (driven by the external stress and also the internal interaction).

\subsection{Temperature-dependent critical Stresses}\label{Sec-T-Tau}

\begin{figure}[!htb]
\centering
\includegraphics[width=0.45\textwidth]{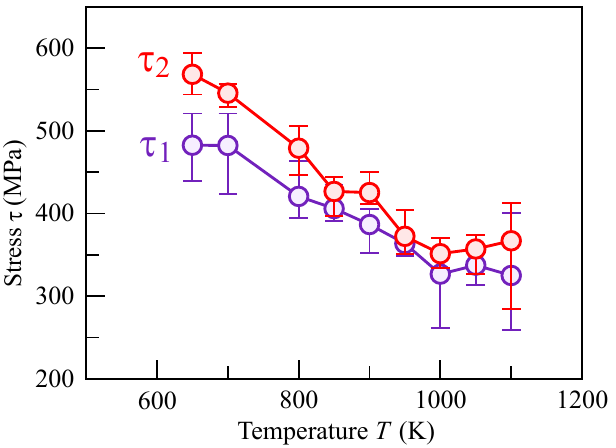}
\caption{Temperature-dependent critical stresses at a shear rate of $\dot{\gamma}=2\times10^{-6}$~ps$^{-1}$.
Error bars are determined from three statistical samples.
% Results from the athermal model~\cite{caillard1996model1} are based on the APB energies obtained using the present MTP (cf. Supplementary Material for details).
}
\label{fig:Tau-T}
\end{figure}

Figure~\ref{fig:Tau-T} shows the temperature-dependence of the stresses required to unlock a KWL. The lower stress $\tau_1$ corresponds to the first cross-slip of the trailing superpartial. The higher stress $\tau_2$ reflects the highest stress required to unlock the KWL, which can correspond for example to the second cross-slip of the trailing superpartial or a dislocation jump (cf.~discussion in Sec.~\ref{sec:unlock}).
For both stresses, a strong decrease with temperature is observed.
In particular, the unlocking stress $\tau_2$ significantly drops by almost 40\% from 568\,MPa at 650\,K to 351\,MPa at 1000\,K, suggesting that in the temperature region of YSA, the unlocking of KWLs has a substantial thermal contribution.
% in addition to the athermal component as was approximated~\cite{caillard1996model1,choi2007modelling}.

% The unlocking stress from the athermal model (cf. Equation~(\ref{seq_un-IKWL-2d}) in Supplementary Material) is displayed as the gray curve in Figure~\ref{fig:Tau-T}.
% which is for KWLs with a cross-slip distance of $\delta$.
% Note that, according to that model, the stress to unlock incomplete KWLs of $2\delta$ should be of 6.5~MPa higher than the gray curve, which will not influence the following discussion.
% Below 800\,K, the unlocking stress in this work is much higher than that from the analytical model.
% This is due to the fact that, in addition to the driving force originated from the internal interaction (the athermal part), the CSF ribbon strongly hinders the movement of $1/2\left<110\right>$ superpartials on the (100) plane.
% In shear simulations at even lower temperatures, for example at 600\,K, the investigated KWL can hardly be unlocked even with a stress up to 760\,MPa.

Above 1000\,K, the two critical stresses exhibit significant statistical fluctuations.
Within the available statistics, the averaged $\tau_1$ remains almost unchanged, whereas the unlocking stress $\tau_2$ increases slightly with temperature.
The latter is due to the increased cross-slip distances at high temperatures (cf.~Figure~\ref{fig_un-IKWL-1050K}).
% In particular at 1100\,K, a jerky gliding on (100) plane over several atomic planes was observed.
In this temperature region, the weak temperature dependence indicates that the thermal component required to constrict the Shockley partials is rather small and the unlocking process depends mainly on the internal elastic interactions between two superpartials, i.e., the athermal component.
% , which is almost constant.

\subsection{Phenomenological Model}

With the atomistically obtained temperature-dependent unlocking stresses, we derive a phenomenological model to predict the unlocking of KWLs, the essence of which is schematically depicted in~Figure~\ref{fig_IKWL-model}(a).
We decompose the effective stress $\tau_{\mathrm{eff}}$ required to unlock a KWL into a static component $\tau_\text{st}$ that balances the internal interaction and a thermal component $\tau^*$ required to overcome the barrier of the thermally-activated unlocking,
\begin{equation}\label{eq_decompose}
    \tau_{\mathrm{eff}} (T) = \tau_\text{st}(T) + \tau^*(T) = \underbrace{\tau_{\mathrm{at}}-\tau_{\mathrm{soft}}(T)}_{\mathrm{static}~\tau_\text{st}}+\tau^*(T),
\end{equation}
where, as indicated, $\tau_\text{st}$ is further decomposed into an athermal term $\tau_{\mathrm{at}}$ and a softening term $\tau_{\mathrm{soft}}$.
The thermal softening term $\tau_{\mathrm{soft}}$ accounts for, e.g., the reduced elastic constants at elevated temperatures. The frictional force is implicitly taken into account by fitting to the unlocking stresses.

For a thermally-activated process, the energy barrier $\Delta H$ can be expressed as~\cite{kocks1975thermodynamics,nemat2000flow,yin2021atomistic,zotov2021molecular}
\begin{equation}\label{eq_enthalpy}
    \Delta H = H_0\left[ 1- \left(\frac{\tau^*}{\tau^*_{\mathrm{0K}}} \right)^p \right]^q,
\end{equation}
where $H_0$ is the enthalpy when $\tau^*$ is zero and $\tau^*_{\mathrm{0K}}$ is the stress to overcome the barrier at 0\,K;
$p$ and $q$ are activation exponents, for which $p=0.5$ %for long range interactions 
is widely used for dislocations~\cite{zotov2021molecular,nemat2000flow,choi2007modelling} and $q=3/2$ is considered to be general for thermally-activated processes~\cite{li2007mechanics,choi2007modelling}.
%(Note that $p=0.5$ and $q=2$ leads to very similar results for the temperature region of interest.)

The energy barrier $\Delta H$ is related to the strain rate $\dot \gamma $ via~\cite{zotov2021molecular,nemat2000flow}
\begin{equation}\label{eq_gamma}
    \frac{\dot \gamma}{\dot{\gamma}_0} = \exp{\left(-\frac{\Delta H}{k_{\mathrm{B}} T}\right)},
\end{equation}
with $\dot{\gamma}_0$ a pre-factor and $k_\mathrm{B}$ the Boltzmann constant. 
Combining Equations~(\ref{eq_enthalpy}) and~(\ref{eq_gamma}), the thermal part $\tau^*$ can be written as:
\begin{equation}\label{Eq_Kocks}
    \tau^*(T) = \tau^*_{\mathrm{0K}} \left[ 1- \left( - \frac{k_{\mathrm{B}} T}{H_0} \ln{\frac{\dot{\gamma}}{\dot{\gamma}_0}} \right)^{2/3} \right]^2.
\end{equation}
% The term inside of the square brackets should be positive.
With increasing temperature, the term inside of the square brackets decreases from 1 at $T=0$\,K to 0 at a temperature at which the process is fully activated and which we label $T_1$.
With the assumption of a linear temperature dependence of the thermal softening term, i.e., $\tau_{\mathrm{soft}}(T)=\epsilon\cdot T$, the effective stress reads
\begin{equation}\label{Eq_Kocks-modified}
    \tau_{\mathrm{eff}} (T) \:=\: \tau_{\mathrm{at}}-\tau_{\mathrm{soft}}(T) + \tau^*(T) \:=\: \tau_{\mathrm{at}} - \epsilon\cdot T + \tau^*_{\mathrm{0K}} \left[ 1- \left( - \frac{k_{\mathrm{B}} T}{H_0} \ln{\frac{\dot{\gamma}}{\dot{\gamma}_0}} \right)^{2/3} \right]^2,
\end{equation}
% where $\tau_\mathrm{a}$, $\tau^*_\mathrm{0K}$, $H_0$ , $\dot{\gamma}_0$, and $\epsilon$ are independent on the strain rate $\dot \gamma$.
with the thermal term $\tau^*(T)$ cut off at $T_1$. We fit the unlocking stresses $\tau_{2} $ obtained from MD simulations at different $\dot{\gamma}$ and $T$ (up to 1000\,K) with the derived phenomenological model (Equation~(\ref{Eq_Kocks-modified})) and obtain the values given in Table~\ref{tab_fit}. The resulting curves for the investigated strain rates are plotted in Figure~\ref{fig_IKWL-model}(b).
While the static term $\tau_\text{st}$, shown as the dashed line labelled ``static'', demonstrates a slight temperature dependence, the thermally-activated term $\tau^*(T)$ represented by the distance between the fitted curve and the ``static'' line rapidly increases with decreasing temperatures.

% \begin{equation}
%    \tau_\mathrm{a} = 463.2\ \text{MPa}, \quad \tau^*_\mathrm{0K} = 2469.1\ \text{MPa}, \quad
%     \dot{\gamma}_0  =  5.07\times 10^{-2}\ \text{ps}^{-1}, \quad
%    H_0 = 1.1\ \text{eV} \quad
%     \epsilon = 0.136\ \text{MPa/K}. \label{eq:fitted}
% \end{equation}

% Note that the obtained $\tau_\mathrm{a}$ of 463.2~MPa is close to the predicted unlocking stress from the athermal model at 0~K (433.9~MPa).
% , supporting the validity of the decomposition utilized in Equation~(\ref{eq_decompose}).
% The enthalpy $H_0$ (1.15~eV) is slightly larger than the result (0.89~eV) derived from DFT calculations~\cite{yu2012effect}, which is thought to result from the lower MTP-predicted CSF formation energy.
% This substantiates the accurate prediction of the energetics in the unlocking processes by the utilized MTP.

\begin{table*}[phbt]
\centering
\begin{tabular}{c|c|c|c|c}
\hline
$\tau_{\mathrm{a}}$ (MPa) &  $\tau_0^*$ (MPa) & $\dot{\gamma}_0$ (ps$^{-1}$) & $H_0$ (eV) & $\epsilon$ (MPa/K)  \\
\hline
 463.2 & 1661.2 & $4.88\times 10^{-2}$ & 1.10 & 0.138  \\
\hline
\end{tabular}
\caption{Optimized fitting parameters for the phenomenological model in Equation \eqref{Eq_Kocks-modified}.}
\label{tab_fit}
\end{table*}

%%%%%%%%%%%%%%%%%%%%%%%%%%%%%%%%%%%%%%%%%%%
%%%%%%%%%%%%%%%%%%%%%%%%%%%%%%%%%%%%%%%%%%%
\section{Discussion}

\begin{figure*}[!t]
\centering
\includegraphics[width=\textwidth]{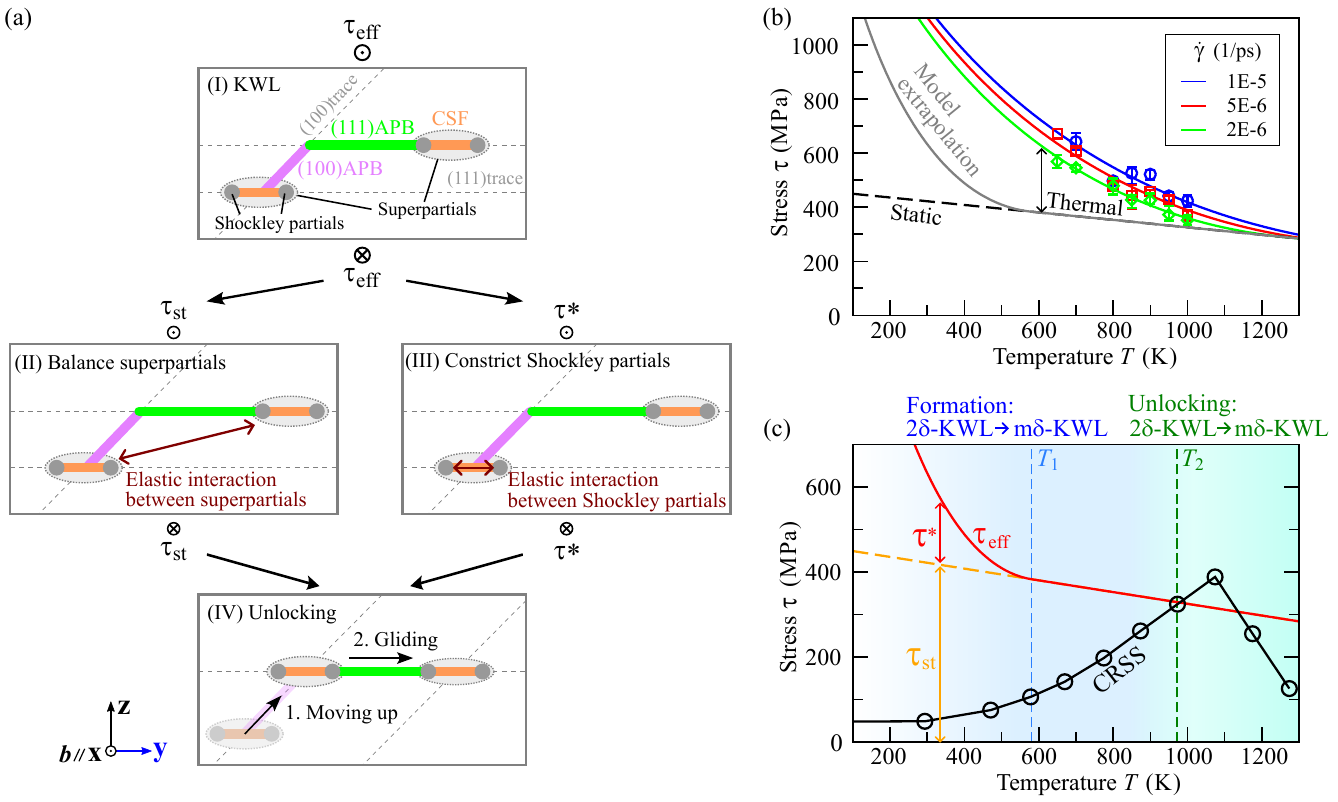}

\caption{Phenomenological model describing the evolution of screw superdislocations in L1$_2$ intermetallics. (a) Schematic diagram for the decomposition of the effective stress to unlock a KWL.
The applied shear stress points into ($\otimes$) or out of ($\odot$) the page.
%The dashed oval represents the core structure of the screw superpartial, which contains a pair of Shockley partials and a CSF ribbon.
For simplicity, the lattice friction acting against dislocation movement is neglected in the illustration.
(b) Effective stresses fitted to Equation~(\ref{Eq_Kocks-modified}) and the model extrapolation to the macroscale (gray curve). The colored symbols denote values from the MD simulations and the corresponding solid curves represent the model fit at the given shear rates.
(c) Comparison between the macroscopic unlocking stress and the critical resolved shear stress (CRSS) (from the $0.2\%$~proof stress)~\cite{golberg1998effect}. 
% $2\delta$-KWLs and $m\delta$-KWLs represent KWLs with a cross-slip distance of $2\delta$ and $m\delta$ ($m>2$), respectively.
}
\label{fig_IKWL-model}
\end{figure*}

The proposed model, derived from atomistic simulations, can be employed to predict the macroscopic unlocking stress and thereby to achieve a better understanding of the YSA. 
% We note that the case studied in this work intuitively reflects macroscopic conditions since YSA is believed to be due to the cross-slip process that exhausts single dislocation~\cite{caillard1996model1,coupeau2020atomic}.
To this end, the macroscopic process is treated as a collection of microscopic unlocking events of single dislocations, each of which can be described by Equation~\eqref{Eq_Kocks-modified} with the here obtained parameters (Table~\ref{tab_fit}).
This is reasonable because the YSA is known to be mainly caused by cross-slip events that lead to an exhaustion of single, mobile dislocations~\cite{caillard1996model1,choi2007modelling,coupeau2020atomic}.
% corresponding to the single dislocation case are independent of the system size.
Therefore, the macroscopic unlocking stress can be obtained by substituting the logarithmic term in Equation~(\ref{Eq_Kocks-modified}) with its  macroscale counterpart $\ln{\left( \dot{\gamma}_\mathrm{m}/\dot{\gamma}_\mathrm{m,0} \right)}$, with $\dot{\gamma}_\mathrm{m}$ and $\dot{\gamma}_{\mathrm{m},0}$ obtained from Ref.~\cite{golberg1998effect} and from Orowan's Equation, respectively (see Supplementary Material for details). The resulting temperature-dependent macroscopic unlocking stress is shown in Figure~\ref{fig_IKWL-model}(b) (gray curve).
The thermal part of the macroscopic unlocking stress is leftwards shifted, i.e., there is a smaller thermal contribution compared to the microscopic curves due to the longer accessible time frames on the macroscale.

The calculated temperature-dependent macroscopic unlocking stress sheds light on the interpretation of the YSA and the accompanying anomalous work-hardening rate, particularly when it is compared with the measured critical resolved shear stress (CRSS) at $0.2\%$ plastic strain for Ni$_3$Al (black curve)~\cite{golberg1998effect}, as plotted in Figure~\ref{fig_IKWL-model}(c). 
Two critical temperatures, $T_1$ and $T_2$, are relevant for the formation and the unlocking of KWLs.
% Three domains dominated by different mechanisms of the KWL structural evolution, separated by two critical temperatures $T_1$ and $T_2$, can be recognized:

\textbf{Formation of KWLs.} 
% The thermally-activated component $\tau^*$ reduces with increasing temperature and becomes zero at $T_1$. Along with the decrease of $\tau^*$, the formation of KWLs driven by internal interactions becomes easier, which leads to an increase of the dislocation exhaustion rate, i.e., to a reduction of the mobile dislocation density, in agreement with experimental studies~\cite{bonneville1996mechanical,kruml2002dislocation}.
The thermally-activated component $\tau^*$ reduces with increasing temperature and becomes zero at $T_1$.
Along with the decrease of $\tau^*$, the formation of KWLs driven by internal interactions  becomes easier due to a reduced cross-slip barrier (without external loading). 
Therefore, mobile dislocations have a rapidly increasing propensity to get exhausted, resulting in an increase of the dislocation exhaustion rate.
This interpretation is in agreement with experimental studies~\cite{bonneville1996mechanical,kruml2002dislocation}.
The CRSS and the work-hardening rate $\theta_{0.2}$ ($\theta$ at $0.2\%$ plastic strain) increase with temperature.
% \item \textbf{CKWL formation ($T_1 < T < T_2$)}
From $T_1$ onward, a Shockley pair bound by a CSF ribbon can easily constrict and, consequently, most screw dislocations transform into $2\delta$-KWLs due to the strong internal driving force.
Since $2\delta$-KWLs with constricted superpartials are only metastable, the superpartials further cross-slip along the (100) plane to form $m\delta$-KWLs with $m>2$ or even complete KWLs, which require higher stresses to unlock.
These types of KWLs have been frequently observed in TEM studies at elevated temperatures~\cite{karnthaler1996influence,rentenberger2001origin}. 
The CRSS and the work-hardening rate $\theta_{0.2}$ increase with temperature at a faster rate.

\textbf{Unlocking of KWLs.} At $T_2$, the CRSS reaches the stress required to unlock $2\delta$-KWLs as predicted by the proposed model. The unlocking of $2\delta$-KWLs suppresses the formation of $m\delta$-KWLs with $m>2$, thus reducing the exhaustion rate and the work hardening rate $\theta_{0.2}$.
The work-hardening rate $\theta_{0.2}$ consequently reaches its maximum around $T_2$.
A similar conclusion was made in Ref.~\cite{kruml1997deformation} where various experimental measurements were carefully analyzed and in Ref.~\cite{conforto2005comparison} where the YSA of three L1$_2$ intermetallics with extreme values of planar-defect formation energies was compared.
The here obtained $T_2 =970$\,K compares well with the experimentally measured value of 800-850\,K at $0.2\%$ plastic strain~\cite{bonneville1996mechanical,conforto2005comparison}. 
% At $T_2$, the unlocking of $2\delta$-KWLs becomes possible, but unlocking $m\delta$-KWLs with $m>2$ remains difficult. 
% The CRSS further increases until a critical, maximal value, at which a statistically significant fraction of $m\delta$-KWLs with $m>2$ can be unlocked and after which various gliding mechanisms become active, e.g., gliding on the (100) plane; APB dragging mechanism~\cite{rentenberger2001origin}.
Upon a further increase of the CRSS, $m\delta$-KWLs with $m>2$ can be unlocked, and at the critical, maximal value of the CRSS, other gliding mechanisms become active, e.g., gliding on the (100) plane~\cite{wang2017understanding} and the APB dragging mechanism~\cite{rentenberger2001origin}.

The proposed phenomenological model does not only explain the  macroscopic YSA of Ni$_3$Al via the evolution of KWLs, but also clarifies the role of the different planar defects on the YSA by the decomposition into a ``static'' and a ``thermal'' component (cf.~Figure~\ref{fig_IKWL-model}). 
The difference in the two types of APBs contributes mostly to the internal elastic interactions between the superpartials, determining the magnitude of the static component (cf.~Equation~\eqref{seq_CS_form}). The CSF energy controls the thermally-activated constriction of the Shockley partials. 
Note that the static component turns into the driving force during the formation of KWLs acting against the thermal constriction.
% A detailed investigation on the formation of KWLs is necessary to complete the understanding of KWLs, which is beyond the scope of this work.
% Furthermore, the starting temperatures of YSA in different L1$_2$ intermetallics are supposed to rely on both the APB-anisotropy and the CSF ribbons, which can also be analyzed by using Equation~\eqref{Eq_Kocks-modified}.

%\subsection{General Applicability of the Phenomenological Model}

The above discussion and the proposed model can be transferred to understand the yield behavior of other L1$_2$ intermetallics, for example, Co-based intermetallics~\cite{suzuki2015l12,chen2022improving} and L1$_2$ strengthened high-entropy alloys~\cite{cao2021novel}.
% The model is constructed in a general way such that it can be applied to other L1$_2$ intermetallics, for example, to Co-based intermetallics~\cite{chen2022improving}.
To apply the model, the system-specific model parameters have to be determined.
For example, $H_0$ can be obtained based on the CSF energy and dislocation interactions~\cite{yu2012effect,vamsi2017yield}; $\tau_\text{at}$ is given by the APB energies and the elastic anisotropic factor~(cf. Equation~\eqref{seq_CS_form}); $\epsilon$ can be evaluated from the temperature dependence of the elastic constants.
With these parameters, $T_1$ and $T_2$ are readily available.
In addition to regulating the formation of KWLs, $T_1$ is also an important factor for determining the activation of gliding systems on $\{100\}$ planes, as previously pointed out for L1$_2$ intermetallics with high CSF energy~\cite{conforto2005comparison}.
In this respect, the proposed model facilitates the establishment of general constitutive laws for describing the yield behavior of L1$_2$ intermetallics.
Tuning YSA for L1$_2$ intermetallics by tailoring the formation energy of planar defects \textit{in silico} is a promising avenue, for example, with the assistance of high-throughput tools~\cite{vamsi2018high}.

In conclusion, KWLs, the origin of YSA in L1$_2$ intermetallics, can be successfully formed and unlocked \textit{in silico} at the atomistic scale by using an \textit{ab initio}-based and physically-informed machine-learning potential.
The simulations show a significant temperature dependence of the unlocking stresses, in contrast to previous athermal predictions.
To describe the unlocking stresses, we have derived a phenomenological model that integrates both the athermal component and the thermally-activated component, and that can be extrapolated to the macroscopic scale. 
By comparing the extrapolated results with experimental values, two critical temperatures are identified, which are of crucial importance to predict the evolution of KWLs.
% $T_1$ demonstrates when the athermal model is valid to use, whereas $T_2$ determines the point when IKWLs start to be unlocked, corresponding further to the temperature when the work-hardening rate peaks.
%Furthermore, through the analysis of atomistic structures, we gained fresh and valuable insights for a better understanding of the dislocation behaviors.
The atomistic simulations predict many cross-slip events, some of which reenact experimental observations, such as the direct formation of $2\delta$-KWLs, the APB-jump, and the dislocation jump. 
The applicability of the here-acquired knowledge and the phenomenological model to other L1$_2$ intermetallics has been elaborated. 

\section{Methods}
\subsection{Moment Tensor Potential}\label{sec_MTP}
The moment descriptor $M_{\mu,\nu}$ in the MTP formalism describes atomic interactions with both radial and angular information according to~\cite{shapeev2016moment}
\begin{equation}
    M_{\mu,\nu}=\sum_{j}{f_\mu (|\vec{r}_{ij}|,z_i,z_j)\underbrace{\vec{r}_{ij}\otimes \dots \otimes \vec{r}_{ij}}_{\nu~\mathrm{times}}}~,
\end{equation}
in which $f_\mu(|\vec{r}_{ij}|,z_i,z_j)$ is the radial function for particle $i$ (type $z_i$ at $\vec{r}_i$) interacting with its neighbor $j$ (type $z_j$ at $\vec{r}_j$); $\mu$ stands for the number of radial functions (depending on the level of the contraction introduced next);
$\vec{r}_{ij}\otimes \dots \otimes \vec{r}_{ij}$ is a tensor of rank $\nu$ representing the angular interaction ($\otimes$ denotes the outer product). 
The scalar contractions of the moments $M_{\mu,\nu}$ give the basis functions $B_\alpha$.
With these basis functions, the local interatomic potential $V(\mathfrak n  _i)$ for atom $i$ with its environment $\mathfrak n_i$ is linearly expanded as
\begin{equation}
    V(\mathfrak n_i)=\sum_\alpha \xi_\alpha B_\alpha (\mathfrak n_i),
\end{equation}
and the energy of the system is then obtained by
\begin{equation}
    E^\mathrm{MTP}=\sum_{i}V(\mathfrak n_i).
\end{equation}
In practice, the number of basis functions is restricted by a degree-like measure, the maximum level $\mathrm{lev}_\mathrm{max}$, and, further, atomic interactions beyond a cutoff radius $R_\mathrm{cut}$ are neglected. To maintain high accuracy at a computationally reasonable number of hyperparameters, a level of $\mathrm{lev}_\mathrm{max}=12$ and a cutoff radius of $R_\mathrm{cut}=5$~\AA\ were chosen.
The weights of the energy and force contributions in the calculation of the loss function were set equal to 1.0, 0.01 \AA$^2$, respectively, while stresses were not considered for training.

\textit{Ab initio} molecular dynamics (AIMD) entering the MTP fitting dataset was performed for all investigated structures at six different volumes at 1600\,K.
The configurations from the AIMD for each structure served as a preliminary information for training the MTP in the corresponding AL loop.

An AL-loop includes four steps: 1) get/train an MTP; 2) perform classic MD simulations with this MTP to select new configurations; 3) perform DFT calculations for selected configurations and 4) go to step 1).
(For a detailed introduction see Ref.~\cite{gubaev2019accelerating}.)
Classic MD simulations were run for 10 picoseconds with the Langevin thermostat at a timestep of 1 femtosecond.
MD snapshots were evaluated by calculating the extrapolation grade $\gamma_\mathrm{mv}$ according to the maxvol algorithm~\cite{gubaev2019accelerating,novikov2020mlip}. 
The threshold to break an MD simulation was set to $\gamma_\mathrm{mv}^\mathrm{break}=3.0$ and the selection threshold to $\gamma_\mathrm{mv}^\mathrm{select}=1.5$.
An AL-loop was finished if no configuration got selected in step 2), meaning that $\gamma_\mathrm{mv}$ for all MD snapshots was below $1.5$.
%This allows only slight extrapolations and makes sure that the knowledge provided for MTP is restricted as intended.
As an exception, the AL-loop for the surface structures was run until the extrapolation grade was below 3.0.
This is sufficient for the surface structures since they are needed only for the free boundary conditions of the shearing simulations and they have negligible influence on the dislocation behavior.

\subsection{Density-functional-theory Calculations}
The DFT calculations were performed by using VASP~\cite{kresse1993ab,kresse1994ab} with potentials based on the projector augmented wave (PAW) method~\cite{blochl1994projector} and within the PBE-GGA approximation \cite{perdew1996generalized} to the exchange-correlation functional. A plane wave cutoff energy of 400\,eV was used.
Regarding the magnetic contribution, which was concluded to have a significant influence on the formation energy of planar defects in particular for the (100)APB~\cite{xu2023APBs}, spin-polarization was considered in all the DFT calculations. 
An MTP trained with a spin-unpolarized dataset was tested and the related discussion is provided in the Supplementary Material (Section~\ref{ssec_NMMTP}).
 
\subsection{Free Energy Calculations}
The Gibbs energy $G(T)$ was obtained via the Legendre transformation of the Helmholtz free energy $F(T)$, which was computed according to
\begin{equation}
    F(T) = E_\mathrm{0K} + F^\mathrm{qh}(T) + F^\mathrm{ah}(T),
\end{equation}
with the total energy at zero-Kelvin $E_\mathrm{0K}$, the quasiharmonic contribution $F^\mathrm{qh}(T)$ and the anharmonic contribution $F^\mathrm{ah}(T)$.
While $F^\mathrm{qh}$ was calculated by using the finite displacement method (pre- and post-processing performed with Phonopy~\cite{phonopy}), $F^\mathrm{ah}$ was obtained by using thermodynamic integration from the quasiharmonic reference to the full vibrational state,
\begin{equation}
    F^\mathrm{ah} = \int_{0}^{1} d \lambda \langle E^\mathrm{vib}-E^\mathrm{qh} \rangle_\lambda,
\end{equation}
with $\lambda$ the coupling factor between the full vibrational state with energy $E^\mathrm{vib}$ and the qh-reference with energy $E^\mathrm{qh}$.

The Gibbs formation energy of the planar defects was then obtained as
\begin{equation}
    \Delta G_\text{form} (T)=\frac{G_\mathrm{defect}(T)-G_\mathrm{bulk}(T)}{A_\mathrm{defect}(T)},
\end{equation}
with the Gibbs energies $G_\mathrm{defect}$ and $G_\mathrm{bulk}$ of the planar defect and bulk supercell, respectively, and with the area of the planar defect $A_\mathrm{defect}$.

\subsection{Molecular Dynamics Simulations}
% An atomistic model containing single screw superdislocation $\left<110\right>$ was established for MD simulations.
ATOMSK~\cite{hirel2015atomsk} was used to generate the initial dislocation configuration according to dislocation theory~\cite{anderson2017theory}.
Specifically, two $1/2\left<1\bar{1}0\right>$ screw superpartials were inserted into the model, bound with a (111)APB region, as shown in Figure~\ref{sfig_cores}(a).
During relaxation, each superpartial dissociated into two Shockley partials, generating a CSF ribbon in between.
The dislocation model was made of 1.5 million atoms with dimensions of $20.3$~nm, $35.1$~nm and $24.8$~nm along the $x$, $y$ and $z$-axis, respectively.
The dislocation line as well as the Burger's vector were placed parallel to the $x$-axis.
% , corresponding to a dislocation density of $1.15\times 10^{15}$~m$^{-2}$.
Periodic boundary conditions were applied along the $x$- and $y$-direction while the boundary conditions along the $z$-direction were shrink-wrapped.

MTP MD simulations were performed with LAMMPS~\cite{thompson2022lammps}. 
To generate an incomplete KWL, the model was first equilibrated at a higher temperature, e.g., 1000\,K, to ensure the activation of the cross-slip process within a reasonable simulation time frame. 
An incomplete KWL with a cross-slip distance of $w=2\delta$ was selected, quenched to 400\,K, and then fully equilibrated at zero pressure.
Before the shearing simulations, three statistical samples initiated with different velocities were heated up to the target temperatures in 10~ps and then fully equilibrated for 20~ps.
All the MD simulations were performed at a timestep of 0.001~ps.

For the shearing simulations, atoms located within a thickness of 9 atomic planes ($\approx1.9$\,nm) on the top and bottom of the box were selected to apply the shear (brown regions in Figure~\ref{sfig_cores}(a)). 
These atoms were treated with flexible boundary conditions to avoid spurious forces on the dislocation~\cite{rodney2007activation}.
The shearing direction was set along the $x$-axis (parallel to the Burger's vector), such that the core of the dislocation was moving along the $y$-axis.

The shearing simulations to unlock the incomplete KWL were performed at a set of strain rates $\dot{\gamma}=1 \times 10^{-5},~5 \times 10^{-6},~2 \times 10^{-6}$~ps$^{-1}$ and temperatures $T=$~650, 700, 800, 850, 900, 950, 1000, 1050, 1100\,K.
Test simulations were also performed with $\dot{\gamma}$ of $1 \times 10^{-6}$~ps$^{-1}$.
% for which the results are displayed in the Supplementary Material (Section~\ref{ssc_conv}).
For each temperature, samples were pre-sheared with $\dot{\gamma} = 1 \times 10^{-4}~\mathrm{ps}^{-1}$ to a pre-strain $\gamma_0$.
Then the production calculations were restarted by continuing shearing from the point of $\gamma_0$ with the same atomic state, e.g., atomic positions, velocities and forces.
The pre-strain $\gamma_0$ was selected such that there was enough time (more than 100\,ps) for equilibration before the occurrence of the first cross-slip, e.g., $\gamma_0 = 0.5\%$ for simulations with $\dot{\gamma}=1 \times 10^{-5},~5 \times 10^{-6}~\mathrm{ps}^{-1}$.
A larger $\gamma_0$ was used in the shearing simulations with $\dot{\gamma}=2 \times 10^{-6}~\mathrm{ps}^{-1}$, to make sure that unlocking is possible within a reasonable computational time (10~ns).
Such a treatment does not only increase the efficiency of the calculations, but also maintains the desired core structure (as illustrated in Figure~\ref{sfig_cores}(c)) for high temperature simulations, e.g., at 1000\,K.
Note that utilizing a high accuracy MTP and performing shearing at a rate of $2\times 10^{-6}$~ps$^{-1}$ (or even slower at $1 \times 10^{-6}$~ps$^{-1}$) for models containing more than one million atoms touches the limits of typical computational resources.

%%%%%%%%%%%%%%%%%%%%%%%%%%%%%%%%%%%%%%%%
\section*{Acknowledgements}
The authors acknowledge fruitful discussions with P. Binkele, N. Zotov, K. Gubaev and P. Kumar.
This work has been funded by the Deutsche Forschungsgemeinschaft (DFG, German Research Foundation) under the Germany's Excellence Strategy - EXC 2075 – 390740016. 
We acknowledge the support by the Stuttgart Center for Simulation Science (SimTech) and the funding from the European Research Council (ERC) under the European Unions Horizon 2020 research and innovation programme (Grant Agreement No. 865855).
We also acknowledge the support by the state of Baden-Württemberg through bwHPC and the German Research Foundation (DFG) through grant no INST 40/575-1 FUGG (JUSTUS 2 cluster).
The shear simulations were performed on the national supercomputer Hawk at the High Performance Computing Center Stuttgart (HLRS) under the grant number H-Embrittlement/44239.

%%%%%%%%%%%%%%%%%%%%%%%%%%%%%%%%%%%%%%%%%%%%%%%%%%%%%%%%%%%%%%%%%%%%%%%%%%%%%%
\bibliographystyle{unsrt}
\bibliography{main}
%%%%%%%%%%%%%%%%%%%%%%%%% Word counting
% end ignoring
%TC:endignore
\end{sloppypar}

\end{document}

\maketitle

\renewcommand\arraystretch{1.2}

\section{Kear-Wilsdorf locks}

\begin{figure}[!hbt]
\centering
\includegraphics[width=\textwidth]{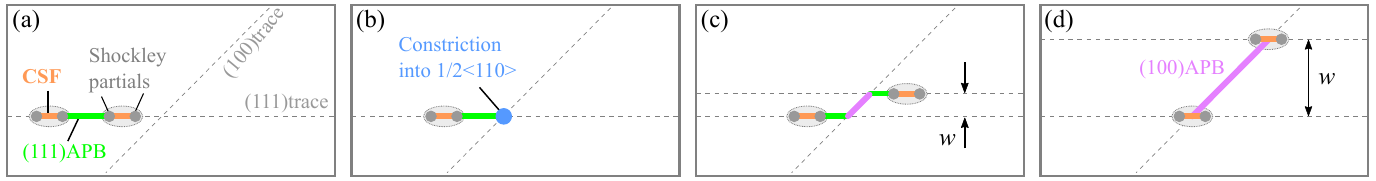}
\caption{Schematic diagram of the formation of a KWL.
% through the cross-slip of a screw superpartial from a $\left(111\right)$ plane to a $\left(100\right)$ plane. 
(a) The core structure of a glissile superdislocation. (b) Constriction of the leading superpartial. (c) An incomplete KWL and (d) a complete KWL. 
}
\label{sfig_KWL}
\end{figure}

Figure~\ref{sfig_KWL} shows schematically the structural evolution of a screw superdislocation in Ni$_3$Al, i.e., the formation of a Kear-Wilsdorf lock (KWL).
The involved planar defects are the antiphase boundaries (APB) on the (100) and (111) planes and the complex stacking fault (CSF).
In (b) a pair of Shockley partials constricts into one superpartial, ready for cross-slipping.
Depending on the cross-slip distance $w$, either an incomplete KWL made of a (100)APB, (111)APB and CSF forms as shown in (c), or a complete KWL develops, for which the superdislocation dissociates into two superpartials only spanning on a (100) plane (no (111)APB) as shown in (d).

\section{Atomistic model}\label{core_conf}
\begin{figure}[!htb]
\centering
\includegraphics[width=0.75\textwidth]{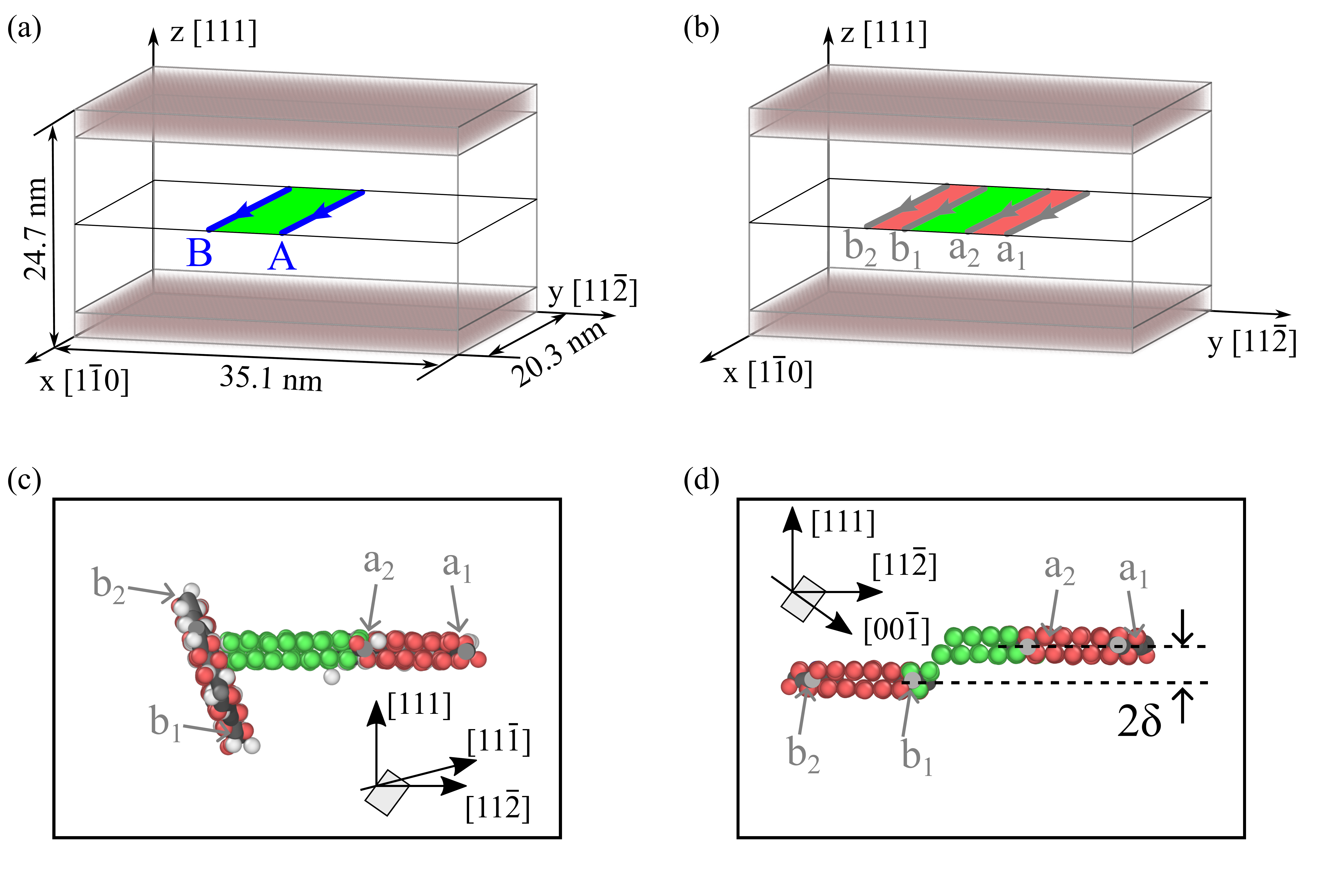}
\caption{Model of the $\left<\bar{1}10\right>$ screw superdislocation in Ni$_3$Al. 
(a) Initial core structure with two superpartials bound with a (111)APB. 
(b) Core structure after relaxation.
Brown regions on the top and bottom of the simulation box are used to perform shearing.
The arrows represent the dislocation line directions.
(c) and (d) Atomistic models for core structures of two incomplete KWLs with a cross-slip distance of $0.5\delta$ and $2\delta$, respectively. 
(c) corresponds to the core configuration described by Equation~\eqref{S3} and (d) to Equation~\eqref{S2}. 
Structural analysis is performed by using OVITO~\cite{stukowski2009visualization} with the DXA algorithm~\cite{stukowski2010extracting}.
Only atoms in the dislocation core structure are shown.
Green and red atoms represent the atoms in the APB and CSF regions.
}
\label{sfig_cores}
\end{figure}

The atomistic model of the core configuration of a screw superdislocation is shown in Figure~\ref{sfig_cores}.
The superdislocation dissociates according to 
\begin{eqnarray}\label{EQ_disl_diss1}
% \begin{aligned}
\left[\bar{1}10\right] \quad
    \rightarrow & \underbrace{\frac{1}{2}\left[\bar{1}10\right]}_\mathrm{A} \qquad\qquad +\quad \text{APB}\quad+\qquad\qquad \underbrace{\frac{1}{2}\left[\bar{1}10\right]}_\mathrm{B} \label{S1}\\ 
    \rightarrow & \underbrace{\frac{1}{6}\left[\bar{1}2\bar{1}\right] +\text{CSF}+\frac{1}{6}\left[\bar{2}11\right]}_{\text{a}_1~\text{and}~\text{a}_2~\text{on a}~(111)~\text{plane}} \quad+\quad \text{APB}\quad+\quad\underbrace{\frac{1}{6}\left[\bar{1}2\bar{1}\right]+\text{CSF}+\frac{1}{6}\left[\bar{2}11\right]}_{\text{b}_1~\text{and}~\text{b}_2~\text{on a}~(111)~\text{plane}} \label{S2}\\
    \rightarrow & \underbrace{\frac{1}{6}\left[\bar{1}2\bar{1}\right] +\text{CSF}+\frac{1}{6}\left[\bar{2}11\right]}_{\text{a}_1~\text{and}~\text{a}_2~\text{on a}~(111)~\text{plane}} \quad+\quad \text{APB}\quad+\quad \underbrace{\frac{1}{6}\left[\bar{2}1\bar{1}\right]+\text{CSF}+\frac{1}{6}\left[\bar{1}21\right]}_{\text{b}_1~\text{and}~\text{b}_2~\text{on a}~(11\bar{1})~\text{plane}} \label{S3}.
% \end{aligned}
\end{eqnarray}
Equation~\eqref{S1} expresses the dissociation of the superdislocation into two superpartials (blue lines labelled A and B in Figure~\ref{sfig_cores}(a)). Each of the superpartials further dissociates into two Shockley partials (green lines a$_1$, a$_2$, b$_1$ and b$_2$ in Figure~\ref{sfig_cores}(b)) according to Equation~\eqref{S2} after relaxation. 
The green and red region represent the (111)APB and the CSF, respectively.
The core configuration described by~Equation~\eqref{S3} is the intermediate state when superpartials cross-slip onto a $\left(\bar{1}\bar{1}1\right)$ plane (equivalent to the $(11\bar{1})$ plane).

% \begin{eqnarray}\label{EQ_disl_diss1}
% % \begin{aligned}
% \left[\bar{1}01\right] \quad
%     \rightarrow & \underbrace{\frac{1}{2}\left[\bar{1}01\right]}_{A} \qquad\qquad +\quad \text{APB}\quad+\qquad\qquad \underbrace{\frac{1}{2}\left[\bar{1}01\right]}_{B} \label{S1}\\ 
%     \rightarrow & \underbrace{\frac{1}{6}\left[\bar{1}12\right] +\text{CSF}+\frac{1}{6}\left[\bar{2}\bar{1}1\right]}_{a_1~\text{and}~a_2~\text{on a}~(111)~\text{plane}} \quad+\quad \text{APB}\quad+\quad\underbrace{\frac{1}{6}\left[\bar{1}12\right]+\text{CSF}+\frac{1}{6}\left[\bar{2}\bar{1}1\right]}_{b_1~\text{and}~b_2~\text{on a}~(111)~\text{plane}} \label{S2}\\
%     \rightarrow & \underbrace{\frac{1}{6}\left[\bar{1}12\right] +\text{CSF}+\frac{1}{6}\left[\bar{2}\bar{1}1\right]}_{a_1~\text{and}~a_2~\text{on a}~(111)~\text{plane}} \quad+\quad \text{APB}\quad+\quad \underbrace{\frac{1}{6}\left[\bar{1}\bar{1}2\right]+\text{CSF}+\frac{1}{6}\left[\bar{2}11\right]}_{b_1~\text{and}~b_2~\text{on a}~(\bar{1}\bar{1}1)~\text{plane}} \label{S3}
% % \end{aligned}
% \end{eqnarray}

\begin{figure}[!htb]
\centering
\includegraphics[width=0.5\textwidth]{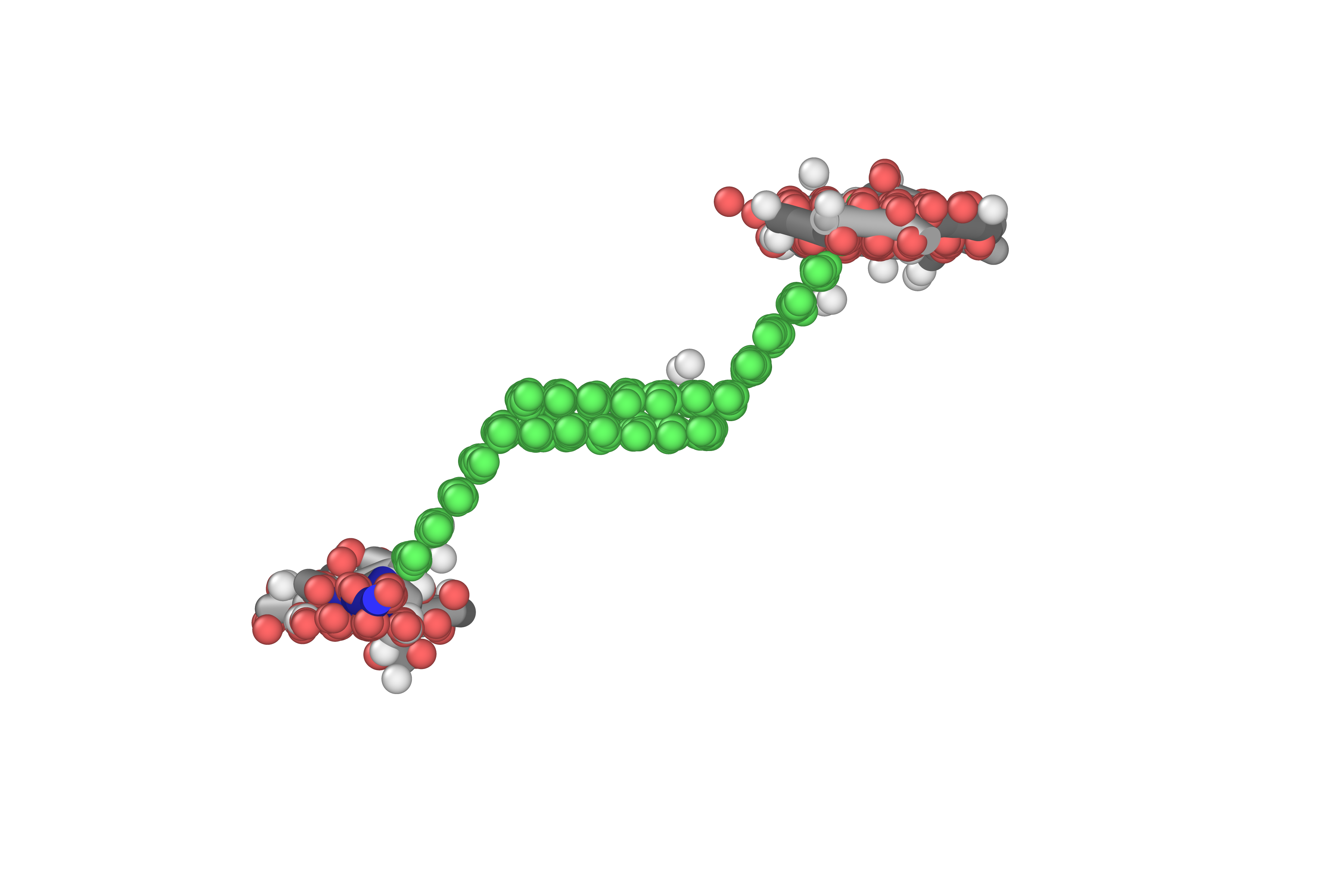}
\caption{Core structure of the $\left<\bar{1}10\right>$ screw superdislocation in Ni$_3$Al after equilibration without external loading at 1100\,K for 1\,ns. 
Atom colors have the same meaning as in Figure~\ref{sfig_cores}.
}
\label{sfig_KWL_1100K}
\end{figure}

When equilibrating without external loading, the leading and trailing superpartial have the same probability to cross-slip onto the (100) plane, which is implied by Figure~\ref{sfig_KWL_1100K}.

\section{Physically-informed active learning}

% The main idea behind the proposed workflow is to apply the traditional active learning (AL) scheme separately for each specified geometry and then to transfer the obtained MTP to the next geometry.
% Preliminary knowledge for learning new structures is provided by few AIMD configurations.
% Note that although the number of provided AIMD configurations is limited, they are essential to ensure a reliable and stable self-learning procedure for new structures (no dubious extrapolation), eventually accelerating the learning process.
% For metallic systems, a logical way to add new structures is from perfect crystal to low dimensional defects and to high dimensional defects. 
% Here we firstly added APBs and stacking faults since they are the most relevant defects to dislocation core structures.

Table~\ref{stab_mAL} records training set sizes and training errors of each active-learning (AL) step for a Moment Tensor Potential (MTP).
The length of the AL-steps after the bulk geometry AL-step are significantly reduced (less configurations selected from AL and less learning iterations).
% The case for (111)APB is exceptional, which is assumed to be due to a rather complex magnetic configuration.
Importantly, when adding a new structure, the training errors change only slightly, indicating the stability of the input MTP potential from the previous step.

% To check whether the final MTP can predict dislocation structures within reasonable extrapolation grade ($\theta < 1.1$), this MTP also went through an active-learning step for single dislocation structure as shown in Figure~\ref{sfig_cores}(b) containing 72000 atoms for 10000 MD steps at 1600\,K.
% The extrapolation grades are all below 1.1, also for cross-slipped configurations, indicating the ability to predict the complex dislocation behaviors for Ni$_3$Al.

\begin{table}[htb]
    \centering
    \begin{threeparttable}
    \begin{tabular}{c|cc|cc|c}
    \hline
     \multirow{2}{*}{Geometries}& \multicolumn{2}{c|}{\# of selected configurations}  & \multicolumn{2}{c|}{RMSE} & \# of learning \\ \cline{2-5} 
      & from AIMD  & from AL & Energy (meV/atom) & Force (eV/\AA) &  steps\\
       \hline
            bulk            &  180 & 218  & 0.66 & 0.047 & 7 \\
            (100)APB        &   45 &  76  & 0.93 & 0.050 & 3 \\
            (111)APB        &  167 &  87  & 1.04 & 0.049 & 6 \\
            CSF             &   45 &  21  & 1.14 & 0.050 & 2 \\
            SISF            &   39 &   6  & 1.15 & 0.049 & 1 \\
            Ni-vacancy      &   45 &  42  & 1.16 & 0.050 & 2 \\
            Al-vacancy      &   45 &  44  & 1.38 & 0.050 & 2 \\
            (100)surface\tnote{\dag}  &  45 &  57  & 1.59 & 0.056 & 2 \\
            (111)surface\tnote{*}  &  0 &  0  & 1.59 & 0.056 & 0 \\
        \hline
    \end{tabular}
    \begin{tablenotes}
        \footnotesize
        \item[\dag] Since the dislocation activity does not depend directly on the surface atoms, the criterion for this AL-step is set as $\theta_{\mathrm{select}}<3$.
        \item[*] The MTP trained with the (100)surface configurations can likewise well describe the (111)surface. 

    \end{tablenotes}

\end{threeparttable}
    \caption{Training set sizes and fitting errors (RMSE: root mean square error) during the physically-informed AL scheme.}
    \label{stab_mAL}
\end{table}

\section{Prediction of elastic constants}
The elastic constants of Ni$_3$Al are provided in Table~\ref{stab_elastic}, including results from experimental and DFT studies.
The anisotropic factor $A$, also known as the Zener factor, is~\cite{yoo1986theory} 
\begin{equation}
    A = \frac{2C_{44}}{C_{11}-C_{12}}~.
\end{equation}
The shear moduli on the (111) plane along [11$\Bar{2}$] and [1$\bar{1}$0] direction are~\cite{jahnatek2009shear} 

\begin{equation}
    \mu_{112} = \frac{3C_{44}(C_{11}-C_{12})}{C_{11}-C_{12}+4C_{44}},
\end{equation}

\begin{equation}
    \mu_{110} = \frac{3C_{44}(C_{11}-C_{12})}{2(C_{11}-C_{12}+C_{44})}.
\end{equation}
In general, $C_{11}$ and $C_{12}$ from the utilized MTP agree well with those from other studies, while $C_{44}$ and $A$ are underestimated.
Further, $B$, $\mu$, $\mu_{112}$ and $\mu_{110}$ also show good agreement with previous studies.
% The driving force for the cross-slip process (cf. Equation~(\ref{seq_CS_form}))  depends slightly on the anisotropic factor $A$.

\begin{table*}[phbt]
\centering
\begin{tabular}{c|ccc|cccccc}
\hline
                    &  $C_{11}$ & $C_{12}$  & $C_{44}$  & $B$   & $Y$   & $A$  & $\mu$ & $\mu_{112}$ & $\mu_{110}$ \\
\hline
MTP (this work)     &  231.8    & 150.3     & 106.6     & 177.5 & 191.3 & 2.62 & 72.5  & 51.28       & 69.28       \\
Prikhodko1999-exp.  &  224.5    & 148.6     & 124.4     & 173.9 & 203.1 & 3.28 & 77.8  & 49.39       & 70.71       \\
Zhao2015-DFT        &  232.7    & 154.5     & 123.0     & 180.6 & 204.0 & 3.14 & 77.8  & 50.61       & 71.82       \\
% Luan2018-DFT        &  240.1 & 160.0 & 123.8 & 186.7 & 78.9 & 207.4 &  3.09 & 51.71 \\
Mishin2004          &  236.0    & 154.3     & 127.1     & 181.5 & 210.9 & 3.11 & 80.7  & 52.95       & 74.60       \\
Du2012              &  242.6    & 149.3     & 130.3     & 180.4 & 223.4 & 2.79 & 86.4  & 59.35       & 81.55       \\
\hline
\end{tabular}
\caption{Elastic properties at 0\,K of Ni$_3$Al in units of GPa ($A$ is dimensionless).
$B$, $\mu$, and $Y$ are the bulk modulus, the shear modulus and the Young's modulus, respectively.
Data from literature are Prikhodko1999 at 300\,K~\cite{prikhodko1999temperature}, Zhao2015-DFT~\cite{zhao2015synergistic} and Luan2018-DFT~\cite{luan2018mechanical}. EAM results: Mishin2004~\cite{mishin2004atomistic}, Du2012~\cite{du2012construction}.}
\label{stab_elastic}
\end{table*}

\section{Atomic extrapolation grades}
\begin{figure}[tb]
\centering
\includegraphics[width=\textwidth]{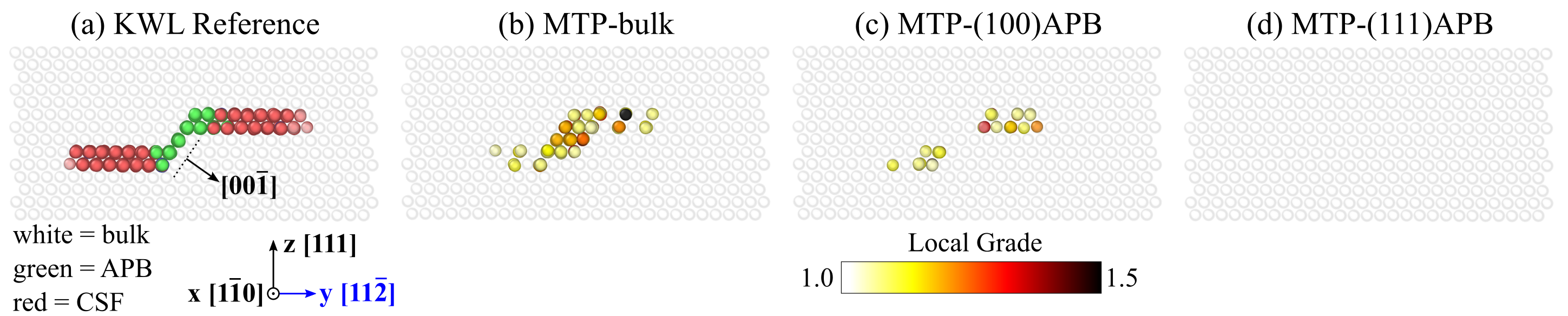}
\caption{
(a) Atomic environment around a KWL.
(b) - (d) Atomic extrapolation grades around the selected KWL environment. 
% The local grades $\theta_\text{local}$ are calculated with MTPs from different stages in the physically informed AL scheme (see text for clear description).
}
\label{sfig_localGrade}
\end{figure}

The atomic extrapolation grades $\theta_\text{local}$ are calculated by using the package \texttt{MLIP-3}~\cite{podryabinkin2023mlip} with an extra function for evaluating atomic neighborhoods.
The functional form of the MTP remains the same as in \texttt{MLIP-2}~\cite{novikov2020mlip} which has been mainly utilized in the present work.
The local grades for the reference KWL structure (Figure~\ref{sfig_localGrade}(a)) were calculated with different MTPs and are shown in Figure~\ref{sfig_localGrade}(b)-(d).
``MTP-bulk'' in (b), ``MTP-100APB'' in (c), and ``MTP-111APB'' in (d) represent MTPs trained with bulk configurations, bulk$+$(100)APB configurations, and bulk$+$(100)APB$+$(111)APB configurations, respectively, following the scheme shown in Figure~1 in the main text.
For the relevant discussion on Figure~\ref{sfig_localGrade}, please see Section 2.1 in the main text.
% Notably, for ``MTP-bulk'', atoms locating near the KWL have larger extrapolation grades.
% Adding (100)APB configurations reduces local grades of atoms in the APB structure in the (100) plane.
% In Figure~\ref{sfig_localGrade}(d) the local grades are all very small, demonstrating that  MTP is able to recognize the atomic environment around a KWL.
%It is thus concluded that the superdislocation core structure can be well captured by the trained MTP following the physically informed active-learning scheme.

\section{Performance of MTP}\label{ssec_NMMTP}
% As shown in Figure~\ref{sfig_ThermProp}, the predicted thermal properties (red curves labeled ``MTP'') show good agreement with experimental and DFT results.
% The deviation in $C_P$ (and $\alpha$ as well) between DFT and MTP originates from the missing entropy contribution of thermal electronic excitations, labeled as ``el'' in Figure~\ref{sfig_ThermProp}(d).

\begin{figure*}[ht]
\centering
\includegraphics[width=\textwidth]{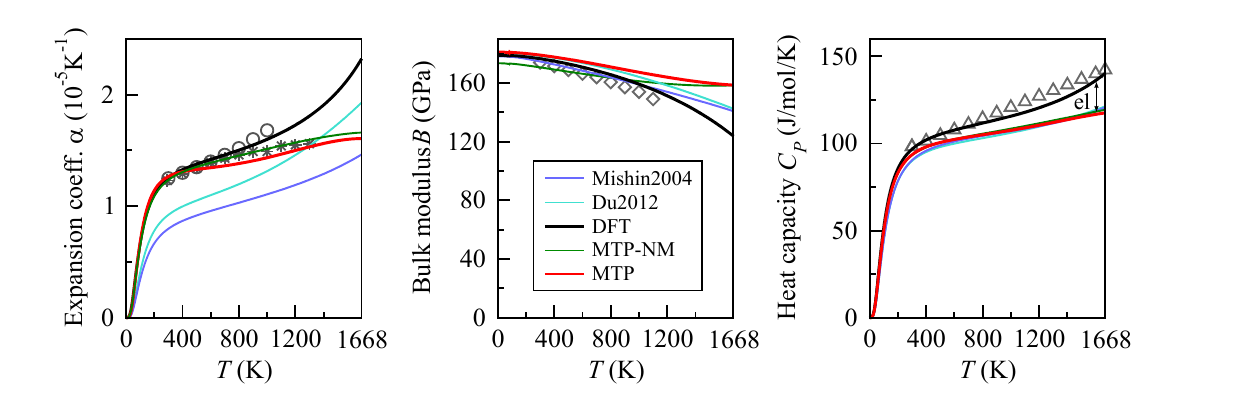}
\caption{Thermal properties of Ni$_3$Al. 
Experimental data: $\alpha$: circles~\cite{touloukian1975thermophysical} and stars~\cite{williams1984physical}; $B$~\cite{prikhodko1999temperature}; $C_p$~\cite{barin1995thermochemical}.
EAM results: Mishin2004~\cite{mishin2004atomistic}, Du2012~\cite{du2012construction}.
}
\label{sfig_ThermProp}
\end{figure*}

\begin{figure*}[ht]
\centering
\includegraphics[width=\textwidth]{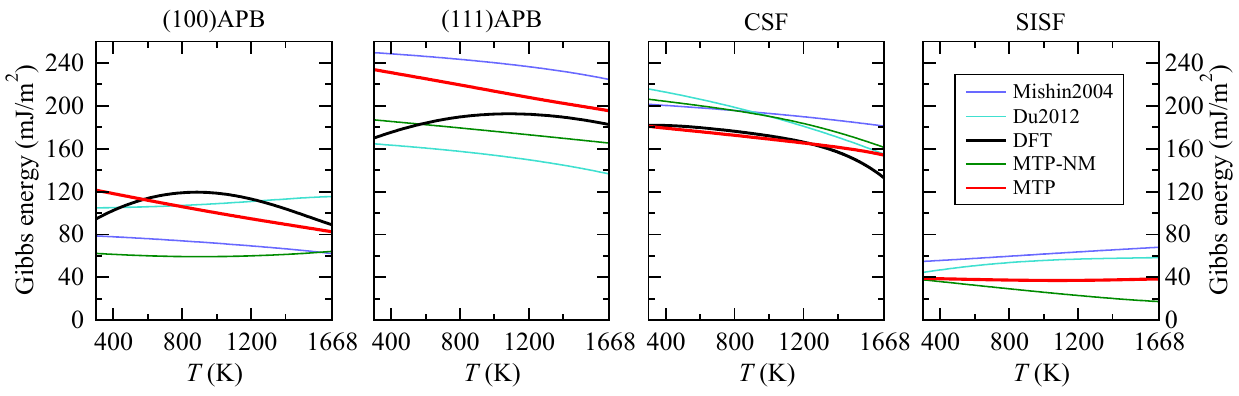}
\caption{Gibbs energy of planar defects in Ni$_3$Al. 
EAM results: Mishin2004~\cite{mishin2004atomistic}, Du2012~\cite{du2012construction}.
}
\label{sfig_GibbsEne}
\end{figure*}

An alternative MTP was trained with a spin-unpolarized DFT dataset following the proposed physically informed active-learning scheme, named as ``MTP-NM'' to distinguish it from the ``MTP'' based on the spin-polarized DFT dataset. 
Its predictive performance is displayed in Figure~\ref{sfig_ThermProp} and~\ref{sfig_GibbsEne}.
From Figure~\ref{sfig_ThermProp}, there is only a slight difference between the thermal properties of the MTPs trained with the spin-polarized and nonpolarized DFT dataset.
However, the Gibbs energies of the planar defects show significant difference between these MTPs, as plotted in Figure~\ref{sfig_GibbsEne}.
Specifically, the predicted Gibbs formation energies   for (100)APB and  for (111)APB from ``MTP-NM'' are smaller than that from ``MTP''.
This is reasonable because magnetism was reported to strongly influence the formation energy for (100)APB and (111)APB~\cite{xu2023APBs}, in particular for (100)APB.
% Indeed, the prediction of $\Gamma_{\mathrm{100}}$ and $\Gamma_{\mathrm{111}}$ from ``MTP-NM'' shows a very good agreement with DFT predictions excluding the magnetic influence, demonstrating the ability of MTPs to reproduce the energy landscape from DFT dataset.
%$\gamma_{\mathrm{CSF}}$ are a bit larger for ``MTP-NM'', which is supposed due to the uncertainty bias of two MTPs. 
Overall, the MTP trained with spin-polarized DFT configurations shows a better agreement with the full DFT results.

% It is stressed that the difference between $\Gamma_{\mathrm{100}}$ and $\Gamma_{\mathrm{111}}$ is much higher for ``MTP-NM'' than for ``MTP'', indicating that a larger stress is needed to unlock KWLs (cf. Figure~\ref{sfig:force}).
% Although the mechanism of dislocation behaviors behind the formation and unlocking of KWLs are supposed to be consistent between two MTPs, from the technical perspective, this challenges the current MD simulations because the simulation time (shear strain) and the cell size (interactions between periodic images) should be increased to properly unlock such a KWL.

% \section{Shear rate dependence of the unlocking stresses}\label{ssc_conv}
% Figure~\ref{sfig_Tau-rate} shows the shear rate dependence of the unlocking stress.
% % As discussed in the main text, $\tau_1$ corresponds to the first upward cross-slip event during unlocking and $\tau_2$ the peak stress.
% % The two stresses are shown in Figure~\ref{sfig_Tau-rate} at different strain rates.

% \begin{figure}[!hbt]
% \centering
% \includegraphics[width=\textwidth]{Tau-rate.pdf}
% \caption{$\tau_1$ and $\tau_2$ at different strain rates $\dot \gamma$.}
% \label{sfig_Tau-rate}
% \end{figure}

% \section{Fitting parameters of the phenomenological model}

% The unlocking stresses at $T \leq 1000$~\,K are used to fit the phenomenological model, because at higher temperatures, KWLs with a cross-slip distance larger than $2\delta$ are often formed.
% This guarantees that the fitted model is for predicting the unlocking process of $2\delta$-KWLs.
% The parameters fitted to Equation~(5) in the main text are listed in Table~\ref{stab_fit}.
% % and the discussion about those parameters are provided in the main text and in the following.

% \begin{table*}[phbt]
% \centering
% \begin{tabular}{c|c|c|c|c}
% \hline
% $\tau_{\mathrm{a}}$ (MPa) &  $\tau_0^*$ (MPa) & $\dot{\gamma}_0$ (ps$^{-1}$) & $H_0$ (eV) & $\epsilon$ (MPa/K)  \\
% \hline
%  463.2 & 1661.2 & $4.88\times 10^{-2}$ & 1.10 & 0.138  \\
% \hline
% \end{tabular}
% \caption{Parameters fitted to Equation~(5) in the main text.}
% \label{stab_fit}
% \end{table*}

\section{Extrapolation of the phenomenological model}
The pre-factor $\dot{\gamma}_{0}$ corresponds to the ideal plastic shear rate when dislocations are mobile, following Orowan's equation
\begin{equation}
    \dot{\gamma}_{0}=\rho b v_0 ,
\end{equation}
with the dislocation density $\rho$, the length of the Burger's vector $b$ and the initial velocity $v_0$. 
For Ni$_3$Al, the initial dislocation density ${\rho}_\mathrm{m}$ was taken to be $1\times 10^{12}$~m$^{-2}$~\cite{kruml2013dislocation}.
$v_0$ can be reasonably approximated as the maximum velocity 1500~m/s from MD simulations~\cite{wakeda2023atomistic}.
Therefore, $\dot{\gamma}_{\mathrm{0,m}}$ is $7.575 \times 10^5$~s$^{-1}$ for a $\left<1\bar{1}0\right>$ superdislocation in Ni$_3$Al with a Burger's vector of 5.05~\AA.
The order of magnitude is comparable with the typically utilized value of $1\times 10^7$~s$^{-1}$ in other studies~\cite{li2021temperature,beyerlein2008dislocation}.
Setting the strain rate $\dot{\gamma}_{\mathrm{m}}=2.3\times 10^{-4}$~s$^{-1}$ as in compression tests~\cite{golberg1998effect}, the model can be extrapolated to the macroscale, by introducing the logarithmic term $\ln{\left( \dot{\gamma}_\text{m}/\dot{\gamma}_\text{0,m} \right)}$.
A similar treatment has been successfully applied in a previous work~\cite{li2021temperature}.

\section{Static analysis based on elastic anisotropy theory}
For a free dislocation (without external loading) spanning on the (111) plane (Figure~\ref{sfig_KWL}(a)), there are internal driving forces that can cause dislocation cross-slip. 
These internal driving forces come from 1) the difference between the energies of the (100)APB and (111)APB~\cite{paidar1984theory} and 2) the elastic anisotropy causing a torque interaction between the two screw superpartials~\cite{yoo1986theory}, which has the form
\begin{equation}\label{seq_CS_form}
    \tau_\mathrm{int} (T) = \frac{1}{b(T)}\left(\frac{\sqrt{3}A}{A+2}\Delta G_{111}(T)-\Delta G_{100}(T)\right).
\end{equation}

\bibliographystyle{unsrt}
\bibliography{supp}